\documentclass[twocolumn,aps,prl,superscriptaddress,showpacs,floatfix,longbibliography]{revtex4-1}
\usepackage{url}
\usepackage{cancel}
\usepackage[colorlinks,linkcolor=blue,citecolor=blue,filecolor=black,urlcolor=blue]{hyperref}
\usepackage{epsfig,graphics}
\usepackage{graphicx}
\usepackage{dcolumn}
\usepackage{bm}
\usepackage[usenames]{color}
\usepackage{amssymb}
\usepackage{amsmath}
\usepackage{multirow}
\usepackage{float}
\usepackage{harpoon}
\usepackage{MnSymbol}
\usepackage{appendix}
\usepackage{color}
\usepackage{hyperref}
\usepackage{cleveref}

\newcommand{\trento}{T$\mathrel{\protect\raisebox{-2.1pt}{R}}$ENTo}

\newcommand{\nch} {N_{\mathrm{ch}}}
\newcommand{\sqrtsnn}{\mbox{$\sqrt{s_{\mathrm{NN}}}$}}
\newcommand{\pT} {p_{\mathrm{T}}}
\newcommand{\lr}[1]{\left\langle #1\right\rangle}

\newcommand{\npart}{N_{\mathrm{part}}}

\newcommand{\ruru}{$^{96}$Ru+$^{96}$Ru}
\newcommand{\zrzr}{$^{96}$Zr+$^{96}$Zr}

\usepackage{CJK}
\hfuzz=\maxdimen
\begin{document}
\title{Separating the impact of nuclear skin and nuclear deformation in high-energy isobar collisions}
\newcommand{\sbu}{Department of Chemistry, Stony Brook University, Stony Brook, NY 11794, USA}
\newcommand{\bnl}{Physics Department, Brookhaven National Laboratory, Upton, NY 11976, USA}
\newcommand{\itp}{Institut f\"ur Theoretische Physik, Universit\"at Heidelberg, Philosophenweg 16, 69120 Heidelberg, Germany}
\author{Jiangyong Jia}\email[Correspond to\ ]{jiangyong.jia@stonybrook.edu}\affiliation{\sbu}\affiliation{\bnl}
\author{Giuliano Giacalone}\email[Correspond to\ ]{giacalone@thphys.uni-heidelberg.de
}\affiliation{\itp}
\author{Chunjian Zhang}\affiliation{\sbu}
\begin{abstract}
Bulk nuclear structure properties, such as radii and deformations, leave distinct signatures in the final state of relativistic heavy-ion collisions. Isobaric collisions offer an easy route to establish explicit connections between the colliding nuclei's structure and the observable outcomes. Here, we investigate the effects of nuclear skin thickness and nuclear deformations on the elliptic flow ($v_2$) and its fluctuations in high-energy $^{96}$Ru+$^{96}$Ru and $^{96}$Zr+$^{96}$Zr collisions. Our findings reveal that the difference in skin thickness between these isobars only influences the inherent ellipticity of the collision systems, $v_2^{\mathrm{rp}}$. In contrast, differences in nuclear deformations solely impact the fluctuations of $v_2$ around $v_2^{\mathrm{rp}}$. Hence, we have identified a data-driven method to disentangle the effects of nuclear skin and nuclear deformations, marking a significant step towards assessing the consistency of nuclear phenomena across energy scales.
\end{abstract}
\maketitle
The bulk properties of atomic nuclei reflect collective correlations in many-body systems held together by the strong force. Unraveling these properties and their evolution across the Segr\'e chart constitutes a major goal in nuclear physics~\cite{Nazarewicz:2016gyu}. Traditionally, spectroscopic and scattering experiments at low energies have been employed to infer collective features of nuclei~\cite{Garrett:2021kfb,Yang:2022wbl,Frois:1987hk}. However, recent ultra-relativistic collision experiments have demonstrated that the dynamics of these collisions is significantly influenced by such properties~\cite{Shou:2014eya,Goldschmidt:2015kpa,Giacalone:2017dud,Giacalone:2019pca,Giacalone:2020awm,Giacalone:2021uhj,Giacalone:2021udy,Jia:2021wbq,Jia:2021tzt,Bally:2021qys,Jia:2021qyu}. Particularly, the angular distributions of emitted particles in high-energy collisions can be related directly to the shape of the colliding nuclei at the moment of interaction.

This connection stems from the near-ideal fluid behavior of the quark-gluon plasma (QGP) formed in high-energy collisions. In a hydrodynamic framework, anisotropies in the final-state azimuthal particle spectra emerge from spatial anisotropies in the initial conditions of the fluid expansion \cite{Gale:2013da,Heinz:2013th,Romatschke:2017ejr}. Spatial anisotropies are, in turn, sourced by the random positions of nucleons populating the colliding nuclei at the time of interaction. Unlike low-energy scattering experiments, which only provide access to average nuclear charge distributions, high-energy heavy-ion collisions can probe the spatial positions of \textit{all} nucleons on an event-by-event basis, thus capturing multi-nucleon correlations.  Experimental techniques employing multi-particle correlations routinely measure these correlations~\cite{Bilandzic:2010jr,Bilandzic:2013kga,Jia:2017hbm}. The crucial question is to what extent the established knowledge from low-energy nuclear physics can offer a coherent understanding of the phenomena observed at high-energy colliders. This Letter represents a significant step in addressing this question.

Although the influence of nuclear deformation is most pronounced in head-on collisions, it is possible to effectively isolate and study nuclear structure effects across the entire centrality range by comparing two isobaric collision systems~\cite{Giacalone:2021uhj,Jia:2021tzt,Jia:2021qyu}. Isobaric nuclei have the same mass number, ensuring that any differences in observables must originate from differences in their structure, which impact the initial condition and evolution of the QGP. This argument is demonstrated clearly in $^{96}$Ru+$^{96}$Ru and $^{96}$Zr+$^{96}$Zr collisions at the BNL Relativistic Heavy Ion Collider, where ratios of observables between the two systems exhibit substantial and centrality-dependent deviations from unity~\cite{STAR:2021mii}. 

Most models describe the nucleon density within colliding nuclei using a Woods-Saxon (WS) profile,
\small{\begin{align}\label{eq:1}
\rho(r,\theta,\phi)\!&\!\propto\![1\!+\!\exp{[r\!-\!R_0\!\left(1\!+\!\beta_2 Y_2^0(\theta,\phi)\!+\!\beta_3 Y_3^0(\theta,\phi)\right)]/a_0}]^{-1},
\end{align}}\normalsize
incorporating four structure parameters, nuclear skin $a_0$, half-width radius $R_0$, quadrupole deformation $\beta_2$, and octupole deformation $\beta_3$. Model studies have demonstrated that isobar ratios are indeed controlled by differences in these parameters, e.g., $\Delta \beta_2^2 = \beta_{\mathrm{2Ru}}^2-\beta_{\mathrm{2Zr}}^2$ , $\Delta \beta_3^2 = \beta_{\mathrm{3Ru}}^2-\beta_{\mathrm{3Zr}}^2$, $\Delta a_0 =a_{\mathrm{0Ru}}-a_{\mathrm{0Zr}}$ and $\Delta R_0 =R_{\mathrm{0Ru}}-R_{\mathrm{0Zr}}$~\cite{Zhang:2021kxj}. 

Experimentally, many observables have been found to exhibit sensitivity to nuclear profile parameters, such as the mean transverse momentum $\pT$~\cite{Xu:2021uar}, its fluctuations~\cite{Jia:2021qyu}, the spectator neutron number~\cite{Liu:2022kvz}, flow vector correlations~\cite{Zhao:2022uhl,Jia:2022qrq}, and shape-size correlations~\cite{Giacalone:2019pca,Bally:2021qys,Jia:2021wbq}.  In this Letter, the focus is on the elliptic flow coefficient $V_2=v_2 e^{2i\Psi_2}$, which characterizes the quadrupole modulation of particles in the direction $\Psi_2$ with an amplitude $v_2$. $V_2$ emerges as a hydrodynamic response to the elliptical shape of the region of overlap between colliding nuclei. The ratio of $v_2$ between \ruru{} and \zrzr{} collisions exhibits a complex non-monotonic centrality dependence \cite{STAR:2021mii}, which can be explained as a combined effect of the four WS parameters in Eq.~\ref{eq:1}~\cite{Jia:2021oyt}. We show that the impact of the deformations parameters ($\beta_{2}$ and $\beta_3$) can be disentangled from that of the radial profile parameters ($a_0$ and $R_0$), and highlight the implications of this finding.

\begin{figure}[t]
\includegraphics[width=0.9\linewidth]{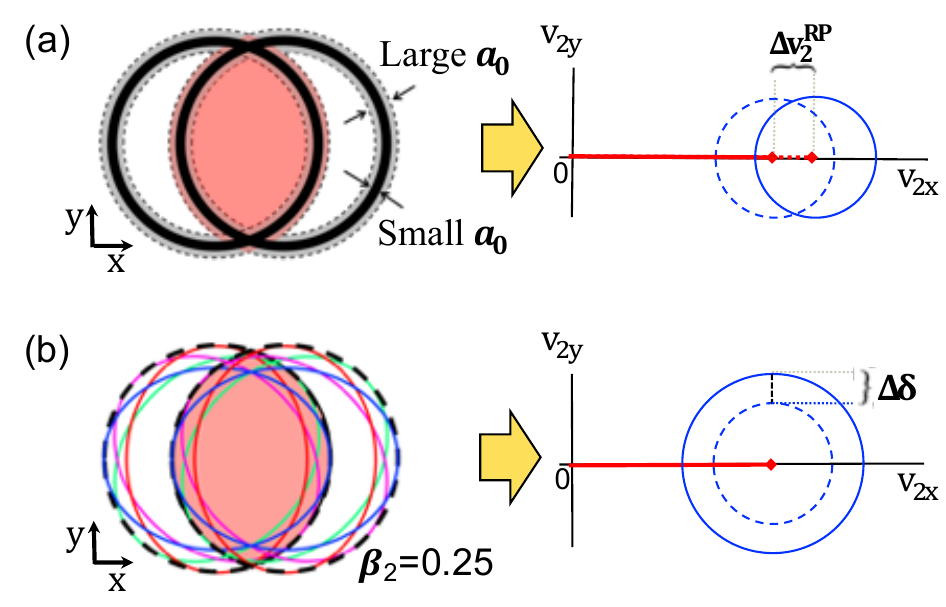}
\vspace*{-.3cm}
\caption{\label{fig:1} (a)  A schematic representation of a collision of spherical nuclei with different choices of their skin thickness, $a_0$. The distribution $(v_{2x},v_{2y})$, denoted by blue circles, has a non-zero mean value along the $x$-direction, $\lr{v_{2x}}=v_2^{\rm rp}$ indicated by red squares at the center of circles, while the variance of the distribution, corresponding to the radius of the circles, is the same along $x$ and $y$. A larger skin (dashed lines) smears the elliptical shape of the QGP, resulting in a reduction of $v_2^{\rm rp}$. (b) Collisions of deformed nuclei with random orientations (four for each nucleus labeled by colored lines) would lead to an increase in the width of the distribution, denoted by $\delta$, relative to collisions of spherical nuclei.}
\end{figure}

We begin with Fig.~\ref{fig:1}, where we represent the plane transverse to the collision axis with Cartesian coordinates, with the $x$ direction aligned with the impact parameter direction. For events at a given centrality, the joint distribution of the real and imaginary parts of $V_2$, $(v_{2x},v_{2y})$, approximately follows a two-dimensional Gaussian distribution \cite{Voloshin:2007pc}
\begin{align}\label{eq:2} 
p(v_{2\mathrm{x}},v_{2\mathrm{y}}) = \frac{1}{\pi\delta^2} \exp \biggl [ - \frac{(v_{2\mathrm{x}}-v_2^{\mathrm{rp}})^2+v_{2\mathrm{y}}^2}{\delta^2} \biggr ].
\end{align}
The displacement along the $x$-axis, $v_2^{\mathrm{rp}}$, corresponds to the reaction plane flow associated with the average elliptic geometry, whereas the fluctuation, $\delta$, represents the variance of elliptic flow due to fluctuations in the positions of the colliding nucleons. Our argument is that changes in the radial profile of the nucleus via $a_0$ or $R_0$ modify $v_2^{\mathrm{rp}}$, while having little impact on the flow fluctuations (Fig.~\ref{fig:1}a). Conversely, in the presence of nuclear deformations, the random orientation of the colliding nuclei results in an increase in $\delta$, with little effect on $v_2^{\mathrm{rp}}$ (Fig.~\ref{fig:1}b)\footnote{Note that changes in nuclear structure also affect the distribution $p(\npart)$, such that the events with the same $\npart$ correspond to slightly different centralities and vice versa~\cite{Jia:2022iji}. This secondary effect introduces a small correlation between $\varepsilon_2^{\mathrm{rp}}$ and $\beta_n$, but it is subleading compared to the one discussed in Fig.~\ref{fig:1}.}. Since $p(V_2)$ is approximately Gaussian, the root-mean-squared elliptic flow is $v_2\{2\}=\sqrt{(v_2^{\mathrm{rp}})^2 + \delta^2}$, while higher-order cumulants of $v_2$ are all identical, $v_2\{4\}=v_2\{6\}=\ldots=v_2\{\infty\}=v_2^{{\mathrm{rp}}}$. In this limit, the fluctuation of $v_2$ can be obtained as
\begin{equation}\label{eq:3} 
\delta^2 = v_2\{2\}^2 - v_2\{4\}^2.
\end{equation}
In the following, we demonstrate our argument regarding the sensitivity of $v_2^{\mathrm{rp}}$ and $\delta$ to the nuclear structure parameters using transport model calculations.

We simulate the dynamics of the QGP using the multi-phase transport model (AMPT)~\cite{Lin:2004en}. Specifically, we use AMPT v2.26t5 in the string-melting mode at $\sqrtsnn=200$~GeV with a partonic cross section of 3.0~$m$b~\cite{Ma:2014pva,Bzdak:2014dia}. This model has been successful in describing the isobar ratios of $v_2$, $v_3$, and $\nch$ measured by the STAR collaboration~\cite{Giacalone:2021udy,Zhang:2021kxj}. We simulate generic isobar collisions, $^{96}$X+$^{96}$X, with five different choices of nuclear structure parameters $\beta_2$, $\beta_3$, $R_0$ and $a_0$, as listed in Table~\ref{tab:1}. This allows us to calculate ratios that isolate the effects of these parameters step-by-step. For example, Case1/Case2 isolates the effect of $\beta_2$, Case1/Case3 includes the effect of $\beta_2$ and $\beta_3$, and so on. We calculate the cumulants of elliptic flow within the multi-particle cumulant framework~\cite{Bilandzic:2010jr,Bilandzic:2013kga} for hadrons with $0.2<\pT<2$ GeV. The two-particle cumulant $v_{2}\{2\}$ is obtained by correlating particles in $0<\eta<2$ with those in $-2<\eta<0$ to suppress short-range correlations that do not arise from the collective expansion of the system~\cite{Jia:2017hbm}. $v_2\{4\}$, which is free from such contributions, is calculated from all particles with $|\eta|<2$. Additionally, we calculate the true $v_2^{\mathrm{rp}}$ from the azimuthal correlation of particles relative to the impact parameter and the true flow fluctuation $\delta_{\mathrm{rp}}$ as $\delta_{\mathrm{rp}}^2 = v_2\{2\}^2 - (v_2^{\mathrm{rp}})^2\;$. The simulated events are binned into classes based on the number of participating nucleons, $\npart$.

\begin{table}[!h]
\centering
\begin{tabular}{|l|cccc|}\hline 
   &\; $R_0$ (fm)\; & \;$a_{0}$ (fm)\;  & $\beta_{2}$ & $\beta_{3}$  \\\hline 
Case1 $^{96}$Ru & 5.09  & 0.46   & 0.162 & 0  \\
Case2          & 5.09  & 0.46   & 0.06  & 0  \\
Case3          & 5.09  & 0.46   & 0.06 & 0.20  \\
Case4          & 5.09  & 0.52   & 0.06 & 0.20  \\
Case5 $^{96}$Zr & 5.02  & 0.52   & 0.06 & 0.20  \\\hline\hline
$\vphantom{\displaystyle\frac{A_A}{A_A}}$ Ratios & \multicolumn{4}{|c|}{$\frac{\mathrm{\textstyle\small Case1}}{\mathrm{ \textstyle\small Case2}}$\;,\; $\frac{\mathrm{\textstyle\small Case1}}{\mathrm{ \textstyle\small Case3}}$\;,\;$\frac{\mathrm{\textstyle\small Case1}}{\mathrm{ \textstyle\small Case4}}$\;,\;$\frac{\mathrm{\textstyle\small Case1}}{\mathrm{ \textstyle\small Case5}}$}\\\hline
\end{tabular}
\caption{\label{tab:1} Structure parameters used in the simulations of \ruru{} and \zrzr{} collisions. Case1 and Case5 represent our full parameterizations of $^{96}$Ru and $^{96}$Zr, respectively.} 
\end{table}

\begin{figure}[t]
\includegraphics[width=0.7\linewidth]{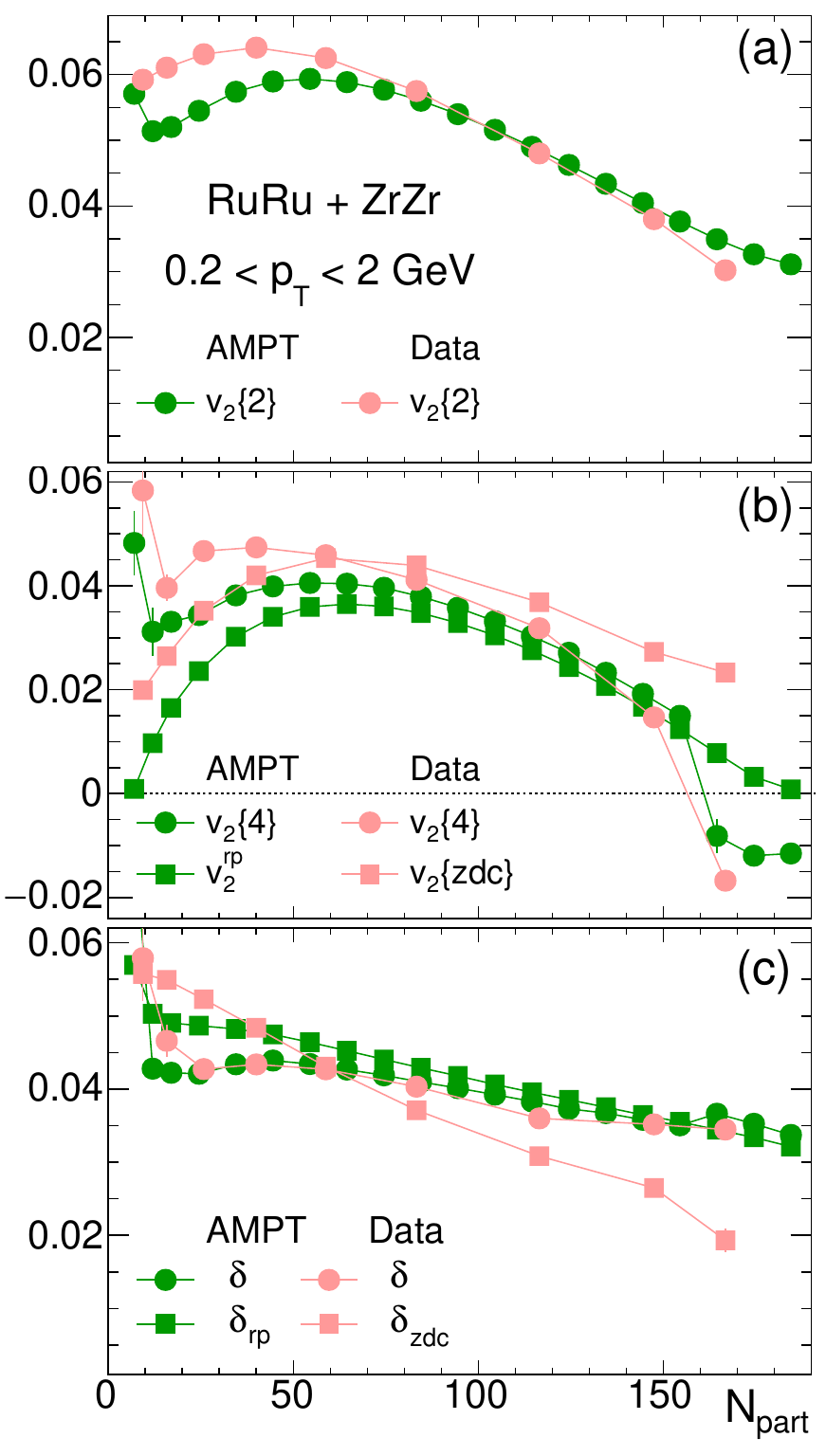}
\vspace*{-.3cm}
\caption{\label{fig:2} Values of $v_2\{2\}$ (a), $v_2\{4\}$ and $v_2^{{\mathrm{rp}}}$ (b), $\delta$ and $\delta_{\mathrm{rp}}$ (c) as a function of $\npart$, averaged between Ru+Ru and Zr+Zr collisions. The AMPT results are compared with the corresponding STAR data from Fig.~23 in Ref.~\cite{STAR:2021mii}. For the STAR data, we approximate $v_2^{{\mathrm{rp}}}$ by $v_{2}\{\mathrm{zdc}\}$, and $\delta_{\mathrm{rp}}$ by $\delta_{\mathrm{zdc}}$, as discussed in the text.}
\end{figure}

In Fig.~\ref{fig:2}, we present our results for $v_2\{2\}$, $v_2\{4\}$, and $\delta$, averaged over \ruru{} and \zrzr{} collisions, which generally agree well with the STAR data. However, we note that the model underpredicts the value of $v_2\{4\}$ in off-central collisions while correctly reproducing the measured $\delta$. This suggests that AMPT has a value of $v_2^{\mathrm{rp}}$ that is too small. This discrepancy may arise from the fact that particle production in AMPT scales with $\npart$, which is known to lead to smaller $v_2^{\mathrm{rp}}$ compared to other models that incorporate proper energy deposition scaling~\cite{Giacalone:2017uqx}. Recent calculations of $v_2\{2\}$ in isobar collisions by Nijs and van der Schee using the \trento{} model do not suffer from this issue~\cite{Nijs:2021kvn}.

The STAR collaboration has also measured an approximation of $v_2^{\mathrm{rp}}$ by correlating particles with spectator neutrons in the zero-degree calorimeters (ZDC), denoted as $v_2\{\mathrm{zdc}\}$.  Figure~\ref{fig:2}(b) shows that $v_2\{\mathrm{zdc}\}$ is smaller than $v_2\{4\}$ in peripheral collisions but is above it towards central collisions. The overall centrality-dependent trend is similar to that of AMPT's $v_2^{\mathrm{rp}}$, indicating that $v_2\{{\mathrm{zdc}}\}$ serves as a good proxy for $v_2^{\mathrm{rp}}$, at least in peripheral and mid-central collisions. In Fig.~\ref{fig:2}(c), we also define the corresponding fluctuation $\delta_{\mathrm{zdc}}$ as $\delta_{\mathrm{zdc}}^2=v_2\{2\}^2 - (v_2\{\mathrm{zdc}\})^2$ and compare the results with the measured $\delta$. They exhibit close agreement in peripheral collisions.

\begin{figure}[t]
\includegraphics[width=\linewidth]{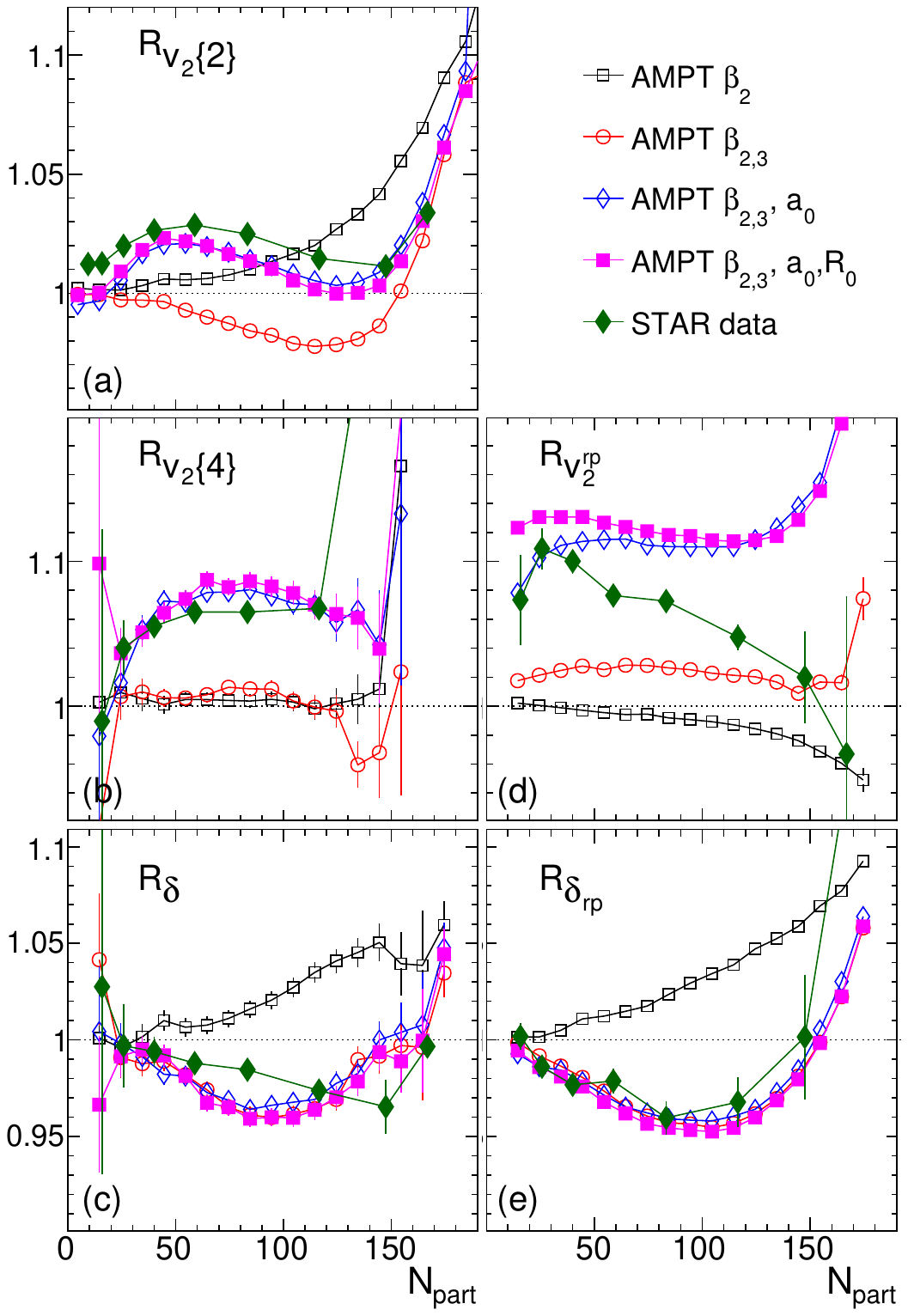}
\vspace*{-.3cm}
\caption{\label{fig:3} Isobar ratios $R_{v_2\{2\}}$ (a), $R_{v_2\{4\}}$ (b), $R_{\delta}$ (c),  $R_{v_2^{\mathrm{rp}}}$ (d), and $R_{\delta_{\mathrm{rp}}}$ (e) plotted as a function of $\npart$.  For clarity and with reference to Tab.~\ref{tab:1}, the curves labeled with ``$\beta_2$'' correspond to Case1/Case2, where the two nuclei differ only by their value of $\beta_2$. The curves labeled with ``$\beta_{2,3}$'' correspond to Case1/Case3, which includes differences in both $\beta_2$ and $\beta_3$.  ``$\beta_{2,3},a_0$'' corresponds to Case1/Case4, adding the difference in $a_0$, while ``$\beta_{2,3},a_0,R_0$" corresponds to Case1/Case5, where all Woods-Saxon parameters are different. The results are compared with STAR data from Fig.~23 of Ref.~\cite{STAR:2021mii}. Note that $v_2^{\mathrm{rp}}$ and $\delta_{\mathrm{rp}}$ cannot be measured directly, and they are approximated by STAR measurements of $v_2\{\mathrm{zdc}\}$ (d) and $\delta_{\mathrm{zdc}}$ (e), respectively.}
\end{figure}

To isolate the effects of nuclear structure, we turn to isobar ratios. For an observable, $\mathcal{O}$, the ratio is calculated at a given $\npart$ as
\begin{align}\label{eq:4}
R_{\mathcal{O}}(\npart)=\frac{\mathcal{O}_{\mathrm{Ru}}(\npart)}{\mathcal{O}_{\mathrm{Zr}}(\npart)}\;.
\end{align}
Figure \ref{fig:3}(a) shows the complex centrality dependence of $R_{v_2\{2\}}$, which arises from both deformation and radial profile parameters. In contrast, $R_{v_2\{4\}}$ in Fig.~\ref{fig:3}(b) is mainly sensitive to $a_0$, whereas $R_{\delta}$ in Fig.~\ref{fig:3}(c) is primarily sensitive to $\beta_2$ and $\beta_3$. Thus, the behavior of $R_{v_2\{2\}}$  can be decomposed into a part that is sensitive to the nuclear skin and a part that is sensitive to the nuclear deformations, supporting the intuition depicted in Fig.~\ref{fig:1}. We establish the following identity,
\begin{align}\label{eq:5}
&R_{v_{2}\{2\}}^2=R_{\delta}^2+(R_{v_{2}\{4\}}^2-R_{\delta}^2)r\;,\; r = v_{2}\{4\}^2/v_{2}\{2\}^2 \\\label{eq:6}
&R_{v_{2}\{2\}} \approx R_{\delta}+ (R_{v_{2}\{4\}}-R_{\delta})r\;,
\end{align}
where the second line is obtained by assuming all ratios are close to unity. It is known that $r\sim0$ in central collisions, and increases to around 0.8 in mid-central collisions. Consequently, the behavior of $R_{v_{2}\{2\}}$ in central collisions is predominantly determined by $R_{\delta}$, while the non-monotonic behavior of $R_{v_{2}\{2\}}$ in mid-central collisions results from the interplay between $R_{v_{2}\{4\}}$ and $R_{\delta}$. This constitutes our main finding. 

The right column of Fig.~\ref{fig:3} presents the ratios of the true intrinsic ellipticity, $R_{v_2^{\mathrm{rp}}}$ in Fig.~\ref{fig:3}(d), and the true flow fluctuation, $R_{\delta_{\mathrm{rp}}}$ in Fig.~\ref{fig:3}(e), obtained from $\delta_{\mathrm{rp}}^2 = v_2\{2\}^2 - (v_2^{\mathrm{rp}})^2$.  We observe that, for a difference in skin thickness of $a_{\mathrm{0Ru}}-a_{\mathrm{0Zr}}=0.06$~fm, the value of $v_2^{{\mathrm{rp}}}$ is enhanced by about 10\% in \ruru{} collisions. The impact of $\beta_n$ on $R_{v_2^{\mathrm{rp}}}$ is relatively minor, as expected. Furthermore, we observe that the values of $v_2^{\mathrm{rp}}$ vary more significantly compared to $v_2\{4\}$ when structure parameters are changed. As a result, $R_{\delta_{\mathrm{rp}}}$ also exhibits a stronger dependence on these parameters than $R_{\delta}$. 

Note that $v_2^{\mathrm{rp}}$ and $\delta_{\mathrm{rp}}$ cannot be measured directly, and therefore the ratios of these observables are approximated by the measured $R_{v_2\{\mathrm{zdc}\}}$ in Fig.~\ref{fig:3}(d) and $R_{\delta_{\mathrm{rp}}}$ in Fig.~\ref{fig:3}(e), respectively. The STAR data agree with AMPT $R_{v_2^{\mathrm{rp}}}$ in peripheral collisions but gradually deviate and loose sensitivity to the WS parameters in more central collisions. This discrepancy could be attributed to a strong decorrelation between the spectator plane and the reaction plane~\cite{ALICE:2022xhd} when the number of spectator neutrons is small. In contrast, STAR $R_{\delta_{\mathrm{zdc}}}$ demonstrates good agreement with AMPT $R_{\delta_{\mathrm{rp}}}$ in Fig.~\ref{fig:3}(e). 

It is worth noting that elliptic flow emerges event-by-event as a response to the initial ellipticity of the system, denoted by $\mathcal{E}_2$. This response follows a linear scaling, $V_2 \propto \mathcal{E}_2$ \cite{Teaney:2010vd}. Therefore, the ratios of observables analyzed in Fig.~\ref{fig:3} can be estimated solely based on knowledge of $\mathcal{E}_2$ and its fluctuations. In the supplemental material, we demonstrate that the observed behaviors in Fig.~\ref{fig:3} largely originate from the initial state.

Our analysis does not rely exclusively on the Gaussian Ansatz in Eq.~\eqref{eq:2} for the distribution of $v_2$. In fact, the fluctuations of $v_2$ are non-Gaussian, especially in peripheral collisions where $v_2^{\mathrm{rp}}$ is large and one becomes sensitive to the bound $v_2<1$ \cite{Giacalone:2016eyu,Bhalerao:2018anl}. It would be interesting to extend this study to higher-order cumulants, $v_2\{4,6,8\}$, and investigate how nuclear structure affects these quantities in isobar collisions~\cite{Giacalone:2018apa}.  In the supplemental material, we provide results for $R_{v_2\{4\}}$, $R_{v_2\{6\}}$ and $R_{v_2\{8\}}$, and also explore the fine splitting of these cumulants in terms of eccentricity fluctuations. Our preliminary findings, limited by AMPT statistics, indicate that there is no apparent separation of nuclear structure effects.

In summary, we have discovered that the nuclear radial profile parameters, i.e., nuclear skin thickness, $a_0$, and half-density radius, $R_0$, predominantly influence the magnitude of $v_2$ along the impact parameter direction captured by $v_2\{4\}$. In contrast, the nuclear deformations, $\beta_n$, primarily affect the fluctuation of elliptic flow, $\delta$. We find that the measured isobar ratio of $v_2\{4\}$ is determined by $a_{\mathrm{0Ru}}-a_{\mathrm{0Zr}}$, while the measured isobar ratio of $\delta$ arises from the interplay between $\beta_{\mathrm{2Ru}}^2-\beta_{\mathrm{2Zr}}^2$ and $\beta_{\mathrm{3Ru}}^2-\beta_{\mathrm{3Zr}}^2$. Our results, combined with the previous finding that the isobar ratio of triangular flow is dominated by $\beta_{\mathrm{3Ru}}^2-\beta_{\mathrm{3Zr}}^2$~\cite{Zhang:2021kxj,Nijs:2021kvn}, provide separate constraints on three key properties of the colliding nuclei: $\Delta a_0$, $\Delta \beta_2^2$, and $\Delta \beta_3^2$.

The skin thickness as a property of the radial structure of nuclei is determined by the frame-independent one-body density of the nuclei. In contrast, deformations are defined in the intrinsic frame of nuclei and can only be captured by two- and many-body densities. Thus, separating skin and deformation effects implies that we have found an experimental method to discern the impact of one-body distribution from that of many-body correlations within nuclei. To our knowledge, such a clean separation of one-body and many-body effects is difficult to achieve in traditional low-energy nuclear structure experiments due to the larger time scales involved. Therefore, our result opens a new opportunity for nuclear structure research based on high-energy nuclear collisions.

On the side of heavy ion physics, our results can aid in the characterization of the QGP from data, which is currently limited by uncertainties in the QGP initial condition~\cite{JETSCAPE:2020mzn,Nijs:2020ors}. Reducing these uncertainties requires improving our understanding of the role of the low-energy structure of nuclei in these processes.  While flow observables are very sensitive to structure parameters between isobars--up to 10\% for two-particle observables~\cite{STAR:2021mii} and even larger for higher-order correlations as seen in Fig.~2 of \cite{Bally:2022vgo}-- the influences of different structure parameters are often entangled for most observables. Our technique, which separates the effects of nuclear radial parameters from nuclear shape parameters, represents a significant step towards refining the initial condition. This reinforces the scientific case for using isobar collisions to elucidate the influence of bulk nuclear structure properties in high-energy collisions, as extensively discussed in Ref.~\cite{Bally:2022vgo}. We hope that it will stimulate further investigations using selected isobar pairs at the LHC~\cite{Citron:2018lsq}.
\bigskip

{\bf Acknowledgements:}  This research of J.J and C.Z is supported by DOE DE-FG02-87ER40331. The research of G.G. is funded by the Deutsche Forschungsgemeinschaft (DFG, German Research Foundation) under Germanys Excellence Strategy EXC2181/1-390900948 (the Heidelberg STRUCTURES Excellence Cluster), within the Collaborative Research Center SFB1225 (ISOQUANT, Project-ID 273811115). We acknowledge Somadutta Bhatta and Jean-Yves Ollitrault for useful discussions.

\section*{Supplemental material}
\subsection{Derivation of Eq.~5}
Equation 5 in the main text is a simple algebric identify. First from Eq. 3 in the text we can write the isobar ratio of $v_{2}\{2\}^2$ as
\begin{align}\nonumber
R_{v_{2}\{2\}^2}&= \frac{\delta^2_{\mathrm{Ru}}}{v_{2}\{2\}^2_{\mathrm{Zr}}}+\frac{v_{2}\{4\}^2_{\mathrm{Ru}}}{v_{2}\{2\}^2_{\mathrm{Zr}}}\\\nonumber
&= R_{\delta^2} \frac{\delta^2_{\mathrm{Zr}}}{v_{2}\{2\}^2_{\mathrm{Zr}}}+R_{v_{2}\{4\}^2} \frac{v_{2}\{4\}^2_{\mathrm{Zr}}}{v_{2}\{2\}^2_{\mathrm{Zr}}}\\\nonumber
&= R_{\delta^2} (1-r) +  R_{v_{2}\{4\}^2} r\\\label{eq:a1}
&= R_{\delta^2} +  (R_{v_{2}\{4\}^2} -R_{\delta^2})r
\end{align}
where $r = (v_{2}\{4\}^2/v_{2}\{2\}^2)_{\mathrm{Zr}}\approx (v_{2}\{4\}^2/v_{2}\{2\}^2)_{\mathrm{Ru}}$. Since  $R_{\mathcal{O}^2} = R_{\mathcal{O}}^2$ is valid for any observable $\mathcal{O}$, Eq.~\ref{eq:a1} above is the same as Eq. 5 in the main text. Equation~6 in the main text is then naturally implied, as long as all the ratios are close to unity.

\subsection{Expectation from isobar ratios of initial state estimators}
\begin{figure}[t]
\includegraphics[width=0.95\linewidth]{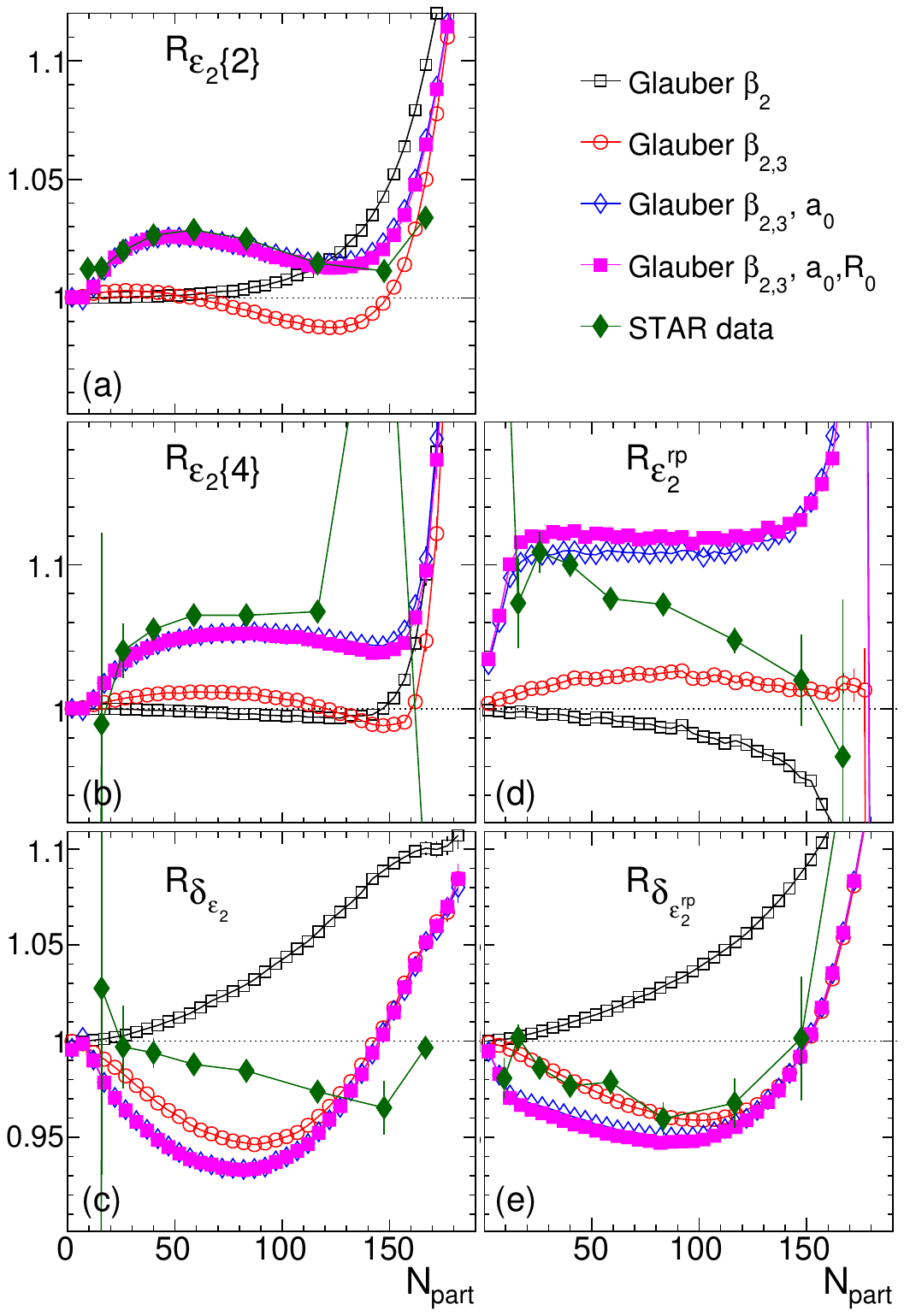}
\vspace*{-.3cm}
\caption{\label{fig:3b} The isobar ratios $R_{\varepsilon_2\{2\}}$ (a), $R_{\varepsilon_2\{4\}}$ (b), $R_{\delta_{\varepsilon}}$ (c),  $R_{\varepsilon_2^{\mathrm{rp}}}$ (e), and $R_{\delta_{\varepsilon^{\mathrm{rp}}}}$ (f) plotted as a function of $\npart$. They are compared with the same STAR data shown in Fig. 3.  Note that $v_2^{\mathrm{rp}}$ and $\delta_{\mathrm{rp}}$ cannot be measured directly, and they are approximated by STAR measurements of $v_2\{\mathrm{zdc}\}$ (d) and $\delta_{\mathrm{zdc}}$ (e), respectively.}
\end{figure}
\begin{figure}[t]
\includegraphics[width=0.9\linewidth]{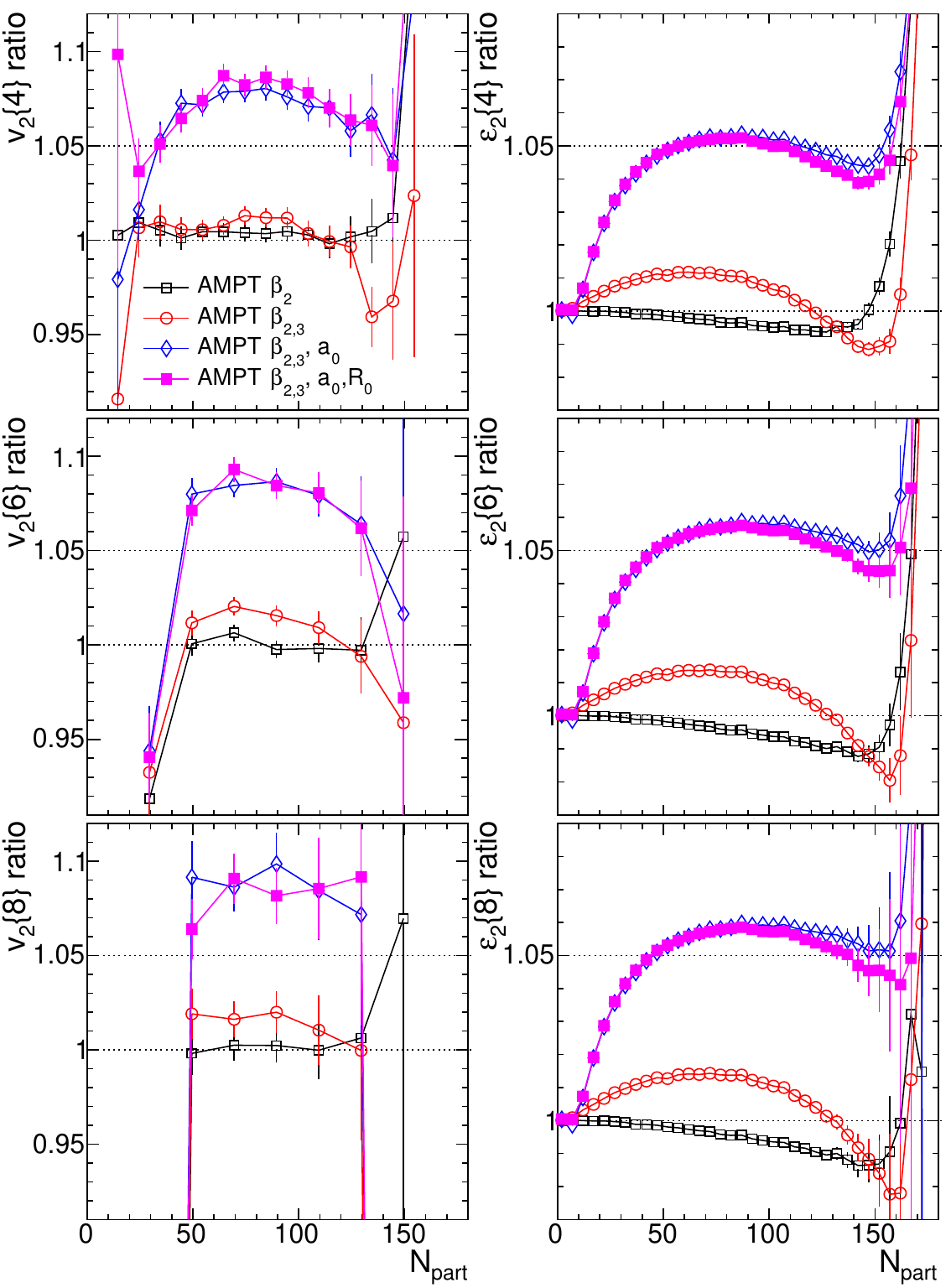}
\vspace*{0cm}
\caption{\label{fig:5} The isobar ratios $R_{v_2\{4\}}$, $R_{v_2\{6\}}$ and $R_{v_2\{8\}}$ (left column) calculated at matching $\npart$ and plotted as a function of $\npart$. The right column shows the ratios calculated based on eccentricity $\varepsilon_2$, $R_{\varepsilon_2\{4\}}$, $R_{\varepsilon_2\{6\}}$ and $R_{\varepsilon_2\{8\}}$. }
\end{figure}
\begin{figure}[t!]
\includegraphics[width=0.9\linewidth]{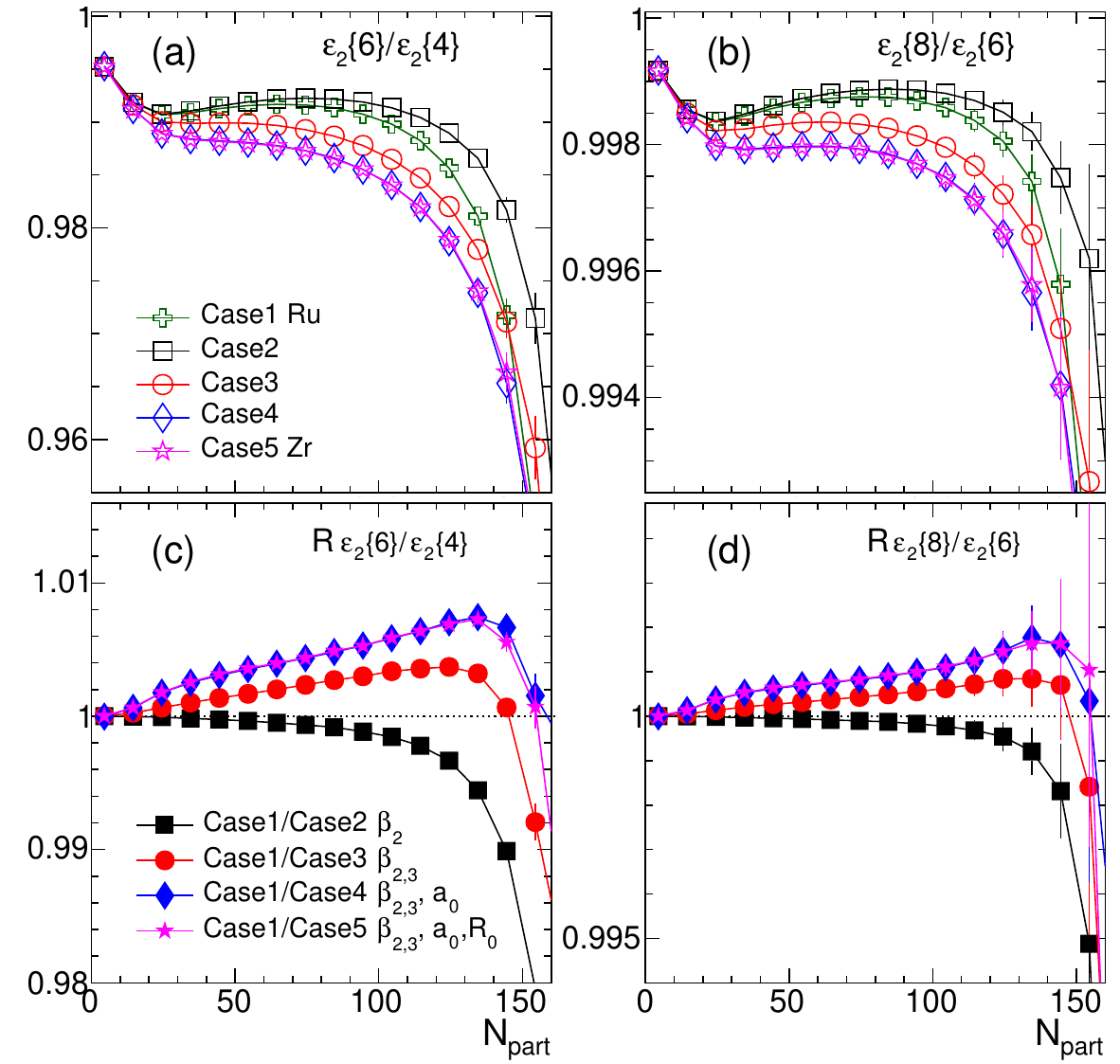}
\caption{\label{fig:6} The ratio of higher-order eccentricity cumulants $\varepsilon_2\{6\}/\varepsilon_2\{4\}$ (a), $\varepsilon_2\{8\}/\varepsilon_2\{6\}$ (b), and corresponding double ratios, $R_{\varepsilon_2\{6\}/\varepsilon_2\{4\}}$ (c), and $R_{\varepsilon_2\{8\}/\varepsilon_2\{6\}}$ (d) to quantify the influence of nuclear structure parameters.}
\end{figure}
A similar analysis is performed for the eccentricity, $\varepsilon_2$, and its fluctuations, where we decompose $\varepsilon_2\{2\}\equiv\sqrt{\lr{\varepsilon_2^2}}$ into a reaction plane component and a fluctuation component. The results are presented in Fig.~\ref{fig:3b}, following the same layout as Fig.3 in the main text. Specifically, we calculate the four-particle cumulant $\varepsilon_2\{4\}$ and the standard reaction plane eccentricity~\cite{Miller:2007ri}, $\varepsilon_2^{\mathrm{rp}}$, as well as the associated fluctuations defined by $\delta_{\varepsilon_2}\equiv \sqrt{\varepsilon_2\{2\}^2-\varepsilon_2\{4\}^2}$ and $\delta_{\varepsilon_2^{\mathrm{rp}}}\equiv \sqrt{\varepsilon_2\{2\}^2-(\varepsilon_2^{\mathrm{rp}})^2}$ to match the corresponding final state quantities. We observe that the isobar ratios of these initial state estimators qualitatively or even quantitatively reproduce the AMPT result in Fig.3, with a few exceptions. Particularly, we note that the values of $R_{\varepsilon_2\{4\}}$ are systematically smaller than $R_{v_2\{4\}}$. However, the values of $R_{\varepsilon_2^{\mathrm{rp}}}$ exhibit quantitative agreement with $R_{v_2^{\mathrm{rp}}}$ in all four cases. The stronger dependence of $R_{\delta_\varepsilon}$ on WS parameters compared to $R_{\delta}$ may be attributed to the role of non-Gaussianities, which are known to be larger in the distribution of $\varepsilon_2$ than in that of $v_2$ due to smearing effects in hydrodynamic expansion. We also observe that $R_{\delta_{\varepsilon_2^{\mathrm{rp}}}}$ agrees quite well with $R_{\delta_{\mathrm{rp}}}$ in Fig.3, as both $\varepsilon_2^{\mathrm{rp}}$ and $v_2^{\mathrm{rp}}$ are defined relative to the impact parameter, and are therefore less affected by non-Gaussianities. However, both $R_{\delta_{\varepsilon_2}}$ and $R_{\delta_{\varepsilon^{\mathrm{rp}}}}$ exhibit some dependence on $a_0$ in peripheral collisions, which is absent in $R_{\delta}$ and $R_{\delta_{\mathrm{rp}}}$ in Fig.3. This is again likely due to a smearing effect in hydrodynamic expansion, which washes out the primordial non-Gaussianities.

Moving on to higher-order fluctuations, we also calculate the ratios of higher-order cumulants, $R_{v_2\{6\}}$ and $R_{v_2\{8\}}$, and compare them to $R_{v_2\{4\}}$. In the Gaussian limit, these ratios should all be identical. However, experimental observations reveal a characteristic fine splitting, $v_2\{4\}\gtrsim v_2\{6\}\gtrsim v_2\{8\}$~\cite{ATLAS:2013xzf,CMS:2017glf,ALICE:2018rtz,ATLAS:2019peb}, reflecting the non-Gaussian nature of the $\varepsilon_2$ distribution \cite{Giacalone:2016eyu,Bhalerao:2018anl}. It is interesting to investigate whether this fine splitting is affected by the differences in nuclear structure. The results are presented in Fig.~\ref{fig:5}. Unfortunately, the statistical precision of our AMPT results does not provide a definitive answer to this question.  On the other hand, the ratios of higher-order cumulants for $\varepsilon_2$ can be calculated with high precision, offering useful guidance. The results in Fig.~\ref{fig:6} bring both good and bad news. The good news is that the cumulant splittings are affected by the nuclear structure parameters. The bad news is that both $\beta_n$ and $a_0$ seem to influence these ratios to a similar extent, reducing the cumulant ratios $\varepsilon_2\{6\}/\varepsilon_2\{4\}$ and $\varepsilon_2\{8\}/\varepsilon_2\{6\}$. It should be noted that the reduction is more pronounced with increasing $\npart$ and may even alter the overall trends of these ratios. While further investigation may be required on the conceptual side, it would still be interesting to explore whether these effects persist in the final state and leave similar imprints on $v_2\{4,6,8\}$, both in simulations and experiments.
\bibliography{deform}{}

\begin{thebibliography}{50}%
\makeatletter
\providecommand \@ifxundefined [1]{%
 \@ifx{#1\undefined}
}%
\providecommand \@ifnum [1]{%
 \ifnum #1\expandafter \@firstoftwo
 \else \expandafter \@secondoftwo
 \fi
}%
\providecommand \@ifx [1]{%
 \ifx #1\expandafter \@firstoftwo
 \else \expandafter \@secondoftwo
 \fi
}%
\providecommand \natexlab [1]{#1}%
\providecommand \enquote  [1]{``#1''}%
\providecommand \bibnamefont  [1]{#1}%
\providecommand \bibfnamefont [1]{#1}%
\providecommand \citenamefont [1]{#1}%
\providecommand \href@noop [0]{\@secondoftwo}%
\providecommand \href [0]{\begingroup \@sanitize@url \@href}%
\providecommand \@href[1]{\@@startlink{#1}\@@href}%
\providecommand \@@href[1]{\endgroup#1\@@endlink}%
\providecommand \@sanitize@url [0]{\catcode `\\12\catcode `\$12\catcode
  `\&12\catcode `\#12\catcode `\^12\catcode `\_12\catcode `\%12\relax}%
\providecommand \@@startlink[1]{}%
\providecommand \@@endlink[0]{}%
\providecommand \url  [0]{\begingroup\@sanitize@url \@url }%
\providecommand \@url [1]{\endgroup\@href {#1}{\urlprefix }}%
\providecommand \urlprefix  [0]{URL }%
\providecommand \Eprint [0]{\href }%
\providecommand \doibase [0]{http://dx.doi.org/}%
\providecommand \selectlanguage [0]{\@gobble}%
\providecommand \bibinfo  [0]{\@secondoftwo}%
\providecommand \bibfield  [0]{\@secondoftwo}%
\providecommand \translation [1]{[#1]}%
\providecommand \BibitemOpen [0]{}%
\providecommand \bibitemStop [0]{}%
\providecommand \bibitemNoStop [0]{.\EOS\space}%
\providecommand \EOS [0]{\spacefactor3000\relax}%
\providecommand \BibitemShut  [1]{\csname bibitem#1\endcsname}%
\let\auto@bib@innerbib\@empty
\bibitem [{\citenamefont {Nazarewicz}(2016)}]{Nazarewicz:2016gyu}%
  \BibitemOpen
  \bibfield  {author} {\bibinfo {author} {\bibfnamefont {Witold}\ \bibnamefont
  {Nazarewicz}},\ }\bibfield  {title} {\enquote {\bibinfo {title} {{Challenges
  in Nuclear Structure Theory}},}\ }\href {\doibase
  10.1088/0954-3899/43/4/044002} {\bibfield  {journal} {\bibinfo  {journal} {J.
  Phys. G}\ }\textbf {\bibinfo {volume} {43}},\ \bibinfo {pages} {044002}
  (\bibinfo {year} {2016})},\ \Eprint {http://arxiv.org/abs/1603.02490}
  {arXiv:1603.02490 [nucl-th]} \BibitemShut {NoStop}%
\bibitem [{\citenamefont {Garrett}\ \emph {et~al.}(2022)\citenamefont
  {Garrett}, \citenamefont {Zieli\'nska},\ and\ \citenamefont
  {Cl\'ement}}]{Garrett:2021kfb}%
  \BibitemOpen
  \bibfield  {author} {\bibinfo {author} {\bibfnamefont {Paul~E.}\ \bibnamefont
  {Garrett}}, \bibinfo {author} {\bibfnamefont {Magda}\ \bibnamefont
  {Zieli\'nska}}, \ and\ \bibinfo {author} {\bibfnamefont {Emmanuel}\
  \bibnamefont {Cl\'ement}},\ }\bibfield  {title} {\enquote {\bibinfo {title}
  {{An experimental view on shape coexistence in nuclei}},}\ }\href {\doibase
  10.1016/j.ppnp.2021.103931} {\bibfield  {journal} {\bibinfo  {journal} {Prog.
  Part. Nucl. Phys.}\ }\textbf {\bibinfo {volume} {124}},\ \bibinfo {pages}
  {103931} (\bibinfo {year} {2022})}\BibitemShut {NoStop}%
\bibitem [{\citenamefont {Yang}\ \emph {et~al.}(2023)\citenamefont {Yang},
  \citenamefont {Wang}, \citenamefont {Wilkins},\ and\ \citenamefont
  {Garcia~Ruiz}}]{Yang:2022wbl}%
  \BibitemOpen
  \bibfield  {author} {\bibinfo {author} {\bibfnamefont {X.~F.}\ \bibnamefont
  {Yang}}, \bibinfo {author} {\bibfnamefont {S.~J.}\ \bibnamefont {Wang}},
  \bibinfo {author} {\bibfnamefont {S.~G.}\ \bibnamefont {Wilkins}}, \ and\
  \bibinfo {author} {\bibfnamefont {R.~F.}\ \bibnamefont {Garcia~Ruiz}},\
  }\bibfield  {title} {\enquote {\bibinfo {title} {{Laser spectroscopy for the
  study of exotic nuclei}},}\ }\href {\doibase 10.1016/j.ppnp.2022.104005}
  {\bibfield  {journal} {\bibinfo  {journal} {Prog. Part. Nucl. Phys.}\
  }\textbf {\bibinfo {volume} {129}},\ \bibinfo {pages} {104005} (\bibinfo
  {year} {2023})},\ \Eprint {http://arxiv.org/abs/2209.15228} {arXiv:2209.15228
  [nucl-ex]} \BibitemShut {NoStop}%
\bibitem [{\citenamefont {Frois}\ and\ \citenamefont
  {Papanicolas}(1987)}]{Frois:1987hk}%
  \BibitemOpen
  \bibfield  {author} {\bibinfo {author} {\bibfnamefont {Bernard}\ \bibnamefont
  {Frois}}\ and\ \bibinfo {author} {\bibfnamefont {Costas~N.}\ \bibnamefont
  {Papanicolas}},\ }\bibfield  {title} {\enquote {\bibinfo {title} {{Electron
  Scattering and Nuclear Structure}},}\ }\href {\doibase
  10.1146/annurev.nucl.37.1.133} {\bibfield  {journal} {\bibinfo  {journal}
  {Ann. Rev. Nucl. Part. Sci.}\ }\textbf {\bibinfo {volume} {37}},\ \bibinfo
  {pages} {133--176} (\bibinfo {year} {1987})}\BibitemShut {NoStop}%
\bibitem [{\citenamefont {Shou}\ \emph {et~al.}(2015)\citenamefont {Shou},
  \citenamefont {Ma}, \citenamefont {Sorensen}, \citenamefont {Tang},
  \citenamefont {Videb\ae{}k},\ and\ \citenamefont {Wang}}]{Shou:2014eya}%
  \BibitemOpen
  \bibfield  {author} {\bibinfo {author} {\bibfnamefont {Q.~Y.}\ \bibnamefont
  {Shou}}, \bibinfo {author} {\bibfnamefont {Y.~G.}\ \bibnamefont {Ma}},
  \bibinfo {author} {\bibfnamefont {P.}~\bibnamefont {Sorensen}}, \bibinfo
  {author} {\bibfnamefont {A.~H.}\ \bibnamefont {Tang}}, \bibinfo {author}
  {\bibfnamefont {F.}~\bibnamefont {Videb\ae{}k}}, \ and\ \bibinfo {author}
  {\bibfnamefont {H.}~\bibnamefont {Wang}},\ }\bibfield  {title} {\enquote
  {\bibinfo {title} {{Parameterization of Deformed Nuclei for Glauber Modeling
  in Relativistic Heavy Ion Collisions}},}\ }\href {\doibase
  10.1016/j.physletb.2015.07.078} {\bibfield  {journal} {\bibinfo  {journal}
  {Phys. Lett. B}\ }\textbf {\bibinfo {volume} {749}},\ \bibinfo {pages}
  {215--220} (\bibinfo {year} {2015})},\ \Eprint
  {http://arxiv.org/abs/1409.8375} {arXiv:1409.8375 [nucl-th]} \BibitemShut
  {NoStop}%
\bibitem [{\citenamefont {Goldschmidt}\ \emph {et~al.}(2015)\citenamefont
  {Goldschmidt}, \citenamefont {Qiu}, \citenamefont {Shen},\ and\ \citenamefont
  {Heinz}}]{Goldschmidt:2015kpa}%
  \BibitemOpen
  \bibfield  {author} {\bibinfo {author} {\bibfnamefont {Andy}\ \bibnamefont
  {Goldschmidt}}, \bibinfo {author} {\bibfnamefont {Zhi}\ \bibnamefont {Qiu}},
  \bibinfo {author} {\bibfnamefont {Chun}\ \bibnamefont {Shen}}, \ and\
  \bibinfo {author} {\bibfnamefont {Ulrich}\ \bibnamefont {Heinz}},\ }\bibfield
   {title} {\enquote {\bibinfo {title} {{Collision geometry and flow in uranium
  + uranium collisions}},}\ }\href {\doibase 10.1103/PhysRevC.92.044903}
  {\bibfield  {journal} {\bibinfo  {journal} {Phys. Rev. C}\ }\textbf {\bibinfo
  {volume} {92}},\ \bibinfo {pages} {044903} (\bibinfo {year} {2015})},\
  \Eprint {http://arxiv.org/abs/1507.03910} {arXiv:1507.03910 [nucl-th]}
  \BibitemShut {NoStop}%
\bibitem [{\citenamefont {Giacalone}\ \emph {et~al.}(2018)\citenamefont
  {Giacalone}, \citenamefont {Noronha-Hostler}, \citenamefont {Luzum},\ and\
  \citenamefont {Ollitrault}}]{Giacalone:2017dud}%
  \BibitemOpen
  \bibfield  {author} {\bibinfo {author} {\bibfnamefont {Giuliano}\
  \bibnamefont {Giacalone}}, \bibinfo {author} {\bibfnamefont {Jacquelyn}\
  \bibnamefont {Noronha-Hostler}}, \bibinfo {author} {\bibfnamefont {Matthew}\
  \bibnamefont {Luzum}}, \ and\ \bibinfo {author} {\bibfnamefont {Jean-Yves}\
  \bibnamefont {Ollitrault}},\ }\bibfield  {title} {\enquote {\bibinfo {title}
  {{Hydrodynamic predictions for 5.44 TeV Xe+Xe collisions}},}\ }\href
  {\doibase 10.1103/PhysRevC.97.034904} {\bibfield  {journal} {\bibinfo
  {journal} {Phys. Rev. C}\ }\textbf {\bibinfo {volume} {97}},\ \bibinfo
  {pages} {034904} (\bibinfo {year} {2018})},\ \Eprint
  {http://arxiv.org/abs/1711.08499} {arXiv:1711.08499 [nucl-th]} \BibitemShut
  {NoStop}%
\bibitem [{\citenamefont {Giacalone}(2020{\natexlab{a}})}]{Giacalone:2019pca}%
  \BibitemOpen
  \bibfield  {author} {\bibinfo {author} {\bibfnamefont {Giuliano}\
  \bibnamefont {Giacalone}},\ }\bibfield  {title} {\enquote {\bibinfo {title}
  {{Observing the deformation of nuclei with relativistic nuclear
  collisions}},}\ }\href {\doibase 10.1103/PhysRevLett.124.202301} {\bibfield
  {journal} {\bibinfo  {journal} {Phys. Rev. Lett.}\ }\textbf {\bibinfo
  {volume} {124}},\ \bibinfo {pages} {202301} (\bibinfo {year}
  {2020}{\natexlab{a}})},\ \Eprint {http://arxiv.org/abs/1910.04673}
  {arXiv:1910.04673 [nucl-th]} \BibitemShut {NoStop}%
\bibitem [{\citenamefont {Giacalone}(2020{\natexlab{b}})}]{Giacalone:2020awm}%
  \BibitemOpen
  \bibfield  {author} {\bibinfo {author} {\bibfnamefont {Giuliano}\
  \bibnamefont {Giacalone}},\ }\bibfield  {title} {\enquote {\bibinfo {title}
  {{Constraining the quadrupole deformation of atomic nuclei with relativistic
  nuclear collisions}},}\ }\href {\doibase 10.1103/PhysRevC.102.024901}
  {\bibfield  {journal} {\bibinfo  {journal} {Phys. Rev. C}\ }\textbf {\bibinfo
  {volume} {102}},\ \bibinfo {pages} {024901} (\bibinfo {year}
  {2020}{\natexlab{b}})},\ \Eprint {http://arxiv.org/abs/2004.14463}
  {arXiv:2004.14463 [nucl-th]} \BibitemShut {NoStop}%
\bibitem [{\citenamefont {Giacalone}\ \emph
  {et~al.}(2021{\natexlab{a}})\citenamefont {Giacalone}, \citenamefont {Jia},\
  and\ \citenamefont {Som\`a}}]{Giacalone:2021uhj}%
  \BibitemOpen
  \bibfield  {author} {\bibinfo {author} {\bibfnamefont {Giuliano}\
  \bibnamefont {Giacalone}}, \bibinfo {author} {\bibfnamefont {Jiangyong}\
  \bibnamefont {Jia}}, \ and\ \bibinfo {author} {\bibfnamefont {Vittorio}\
  \bibnamefont {Som\`a}},\ }\bibfield  {title} {\enquote {\bibinfo {title}
  {{Accessing the shape of atomic nuclei with relativistic collisions of
  isobars}},}\ }\href {\doibase 10.1103/PhysRevC.104.L041903} {\bibfield
  {journal} {\bibinfo  {journal} {Phys. Rev. C}\ }\textbf {\bibinfo {volume}
  {104}},\ \bibinfo {pages} {L041903} (\bibinfo {year} {2021}{\natexlab{a}})},\
  \Eprint {http://arxiv.org/abs/2102.08158} {arXiv:2102.08158 [nucl-th]}
  \BibitemShut {NoStop}%
\bibitem [{\citenamefont {Giacalone}\ \emph
  {et~al.}(2021{\natexlab{b}})\citenamefont {Giacalone}, \citenamefont {Jia},\
  and\ \citenamefont {Zhang}}]{Giacalone:2021udy}%
  \BibitemOpen
  \bibfield  {author} {\bibinfo {author} {\bibfnamefont {Giuliano}\
  \bibnamefont {Giacalone}}, \bibinfo {author} {\bibfnamefont {Jiangyong}\
  \bibnamefont {Jia}}, \ and\ \bibinfo {author} {\bibfnamefont {Chunjian}\
  \bibnamefont {Zhang}},\ }\bibfield  {title} {\enquote {\bibinfo {title}
  {{Impact of Nuclear Deformation on Relativistic Heavy-Ion Collisions:
  Assessing Consistency in Nuclear Physics across Energy Scales}},}\ }\href
  {\doibase 10.1103/PhysRevLett.127.242301} {\bibfield  {journal} {\bibinfo
  {journal} {Phys. Rev. Lett.}\ }\textbf {\bibinfo {volume} {127}},\ \bibinfo
  {pages} {242301} (\bibinfo {year} {2021}{\natexlab{b}})},\ \Eprint
  {http://arxiv.org/abs/2105.01638} {arXiv:2105.01638 [nucl-th]} \BibitemShut
  {NoStop}%
\bibitem [{\citenamefont {Jia}\ \emph {et~al.}(2022{\natexlab{a}})\citenamefont
  {Jia}, \citenamefont {Huang},\ and\ \citenamefont {Zhang}}]{Jia:2021wbq}%
  \BibitemOpen
  \bibfield  {author} {\bibinfo {author} {\bibfnamefont {Jiangyong}\
  \bibnamefont {Jia}}, \bibinfo {author} {\bibfnamefont {Shengli}\ \bibnamefont
  {Huang}}, \ and\ \bibinfo {author} {\bibfnamefont {Chunjian}\ \bibnamefont
  {Zhang}},\ }\bibfield  {title} {\enquote {\bibinfo {title} {{Probing nuclear
  quadrupole deformation from correlation of elliptic flow and transverse
  momentum in heavy ion collisions}},}\ }\href {\doibase
  10.1103/PhysRevC.105.014906} {\bibfield  {journal} {\bibinfo  {journal}
  {Phys. Rev. C}\ }\textbf {\bibinfo {volume} {105}},\ \bibinfo {pages}
  {014906} (\bibinfo {year} {2022}{\natexlab{a}})},\ \Eprint
  {http://arxiv.org/abs/2105.05713} {arXiv:2105.05713 [nucl-th]} \BibitemShut
  {NoStop}%
\bibitem [{\citenamefont {Jia}(2022{\natexlab{a}})}]{Jia:2021tzt}%
  \BibitemOpen
  \bibfield  {author} {\bibinfo {author} {\bibfnamefont {Jiangyong}\
  \bibnamefont {Jia}},\ }\bibfield  {title} {\enquote {\bibinfo {title} {{Shape
  of atomic nuclei in heavy ion collisions}},}\ }\href {\doibase
  10.1103/PhysRevC.105.014905} {\bibfield  {journal} {\bibinfo  {journal}
  {Phys. Rev. C}\ }\textbf {\bibinfo {volume} {105}},\ \bibinfo {pages}
  {014905} (\bibinfo {year} {2022}{\natexlab{a}})},\ \Eprint
  {http://arxiv.org/abs/2106.08768} {arXiv:2106.08768 [nucl-th]} \BibitemShut
  {NoStop}%
\bibitem [{\citenamefont {Bally}\ \emph
  {et~al.}(2022{\natexlab{a}})\citenamefont {Bally}, \citenamefont {Bender},
  \citenamefont {Giacalone},\ and\ \citenamefont {Som\`a}}]{Bally:2021qys}%
  \BibitemOpen
  \bibfield  {author} {\bibinfo {author} {\bibfnamefont {Benjamin}\
  \bibnamefont {Bally}}, \bibinfo {author} {\bibfnamefont {Michael}\
  \bibnamefont {Bender}}, \bibinfo {author} {\bibfnamefont {Giuliano}\
  \bibnamefont {Giacalone}}, \ and\ \bibinfo {author} {\bibfnamefont
  {Vittorio}\ \bibnamefont {Som\`a}},\ }\bibfield  {title} {\enquote {\bibinfo
  {title} {{Evidence of the triaxial structure of $\boldsymbol{^{129}}$Xe at
  the Large Hadron Collider}},}\ }\href {\doibase
  10.1103/PhysRevLett.128.082301} {\bibfield  {journal} {\bibinfo  {journal}
  {Phys. Rev. Lett.}\ }\textbf {\bibinfo {volume} {128}},\ \bibinfo {pages}
  {082301} (\bibinfo {year} {2022}{\natexlab{a}})},\ \Eprint
  {http://arxiv.org/abs/2108.09578} {arXiv:2108.09578 [nucl-th]} \BibitemShut
  {NoStop}%
\bibitem [{\citenamefont {Jia}(2022{\natexlab{b}})}]{Jia:2021qyu}%
  \BibitemOpen
  \bibfield  {author} {\bibinfo {author} {\bibfnamefont {Jiangyong}\
  \bibnamefont {Jia}},\ }\bibfield  {title} {\enquote {\bibinfo {title}
  {{Probing triaxial deformation of atomic nuclei in high-energy heavy ion
  collisions}},}\ }\href {\doibase 10.1103/PhysRevC.105.044905} {\bibfield
  {journal} {\bibinfo  {journal} {Phys. Rev. C}\ }\textbf {\bibinfo {volume}
  {105}},\ \bibinfo {pages} {044905} (\bibinfo {year} {2022}{\natexlab{b}})},\
  \Eprint {http://arxiv.org/abs/2109.00604} {arXiv:2109.00604 [nucl-th]}
  \BibitemShut {NoStop}%
\bibitem [{\citenamefont {Gale}\ \emph {et~al.}(2013)\citenamefont {Gale},
  \citenamefont {Jeon},\ and\ \citenamefont {Schenke}}]{Gale:2013da}%
  \BibitemOpen
  \bibfield  {author} {\bibinfo {author} {\bibfnamefont {Charles}\ \bibnamefont
  {Gale}}, \bibinfo {author} {\bibfnamefont {Sangyong}\ \bibnamefont {Jeon}}, \
  and\ \bibinfo {author} {\bibfnamefont {Bjoern}\ \bibnamefont {Schenke}},\
  }\bibfield  {title} {\enquote {\bibinfo {title} {{Hydrodynamic Modeling of
  Heavy-Ion Collisions}},}\ }\href {\doibase 10.1142/S0217751X13400113}
  {\bibfield  {journal} {\bibinfo  {journal} {Int. J. Mod. Phys.}\ }\textbf
  {\bibinfo {volume} {A28}},\ \bibinfo {pages} {1340011} (\bibinfo {year}
  {2013})},\ \Eprint {http://arxiv.org/abs/1301.5893} {arXiv:1301.5893
  [nucl-th]} \BibitemShut {NoStop}%
\bibitem [{\citenamefont {Heinz}\ and\ \citenamefont
  {Snellings}(2013)}]{Heinz:2013th}%
  \BibitemOpen
  \bibfield  {author} {\bibinfo {author} {\bibfnamefont {Ulrich}\ \bibnamefont
  {Heinz}}\ and\ \bibinfo {author} {\bibfnamefont {Raimond}\ \bibnamefont
  {Snellings}},\ }\bibfield  {title} {\enquote {\bibinfo {title} {{Collective
  flow and viscosity in relativistic heavy-ion collisions}},}\ }\href {\doibase
  10.1146/annurev-nucl-102212-170540} {\bibfield  {journal} {\bibinfo
  {journal} {Ann. Rev. Nucl. Part. Sci.}\ }\textbf {\bibinfo {volume} {63}},\
  \bibinfo {pages} {123--151} (\bibinfo {year} {2013})},\ \Eprint
  {http://arxiv.org/abs/1301.2826} {arXiv:1301.2826 [nucl-th]} \BibitemShut
  {NoStop}%
\bibitem [{\citenamefont {Romatschke}\ and\ \citenamefont
  {Romatschke}(2019)}]{Romatschke:2017ejr}%
  \BibitemOpen
  \bibfield  {author} {\bibinfo {author} {\bibfnamefont {Paul}\ \bibnamefont
  {Romatschke}}\ and\ \bibinfo {author} {\bibfnamefont {Ulrike}\ \bibnamefont
  {Romatschke}},\ }\href {\doibase 10.1017/9781108651998} {\emph {\bibinfo
  {title} {{Relativistic Fluid Dynamics In and Out of Equilibrium}}}},\
  Cambridge Monographs on Mathematical Physics\ (\bibinfo  {publisher}
  {Cambridge University Press},\ \bibinfo {year} {2019})\ \Eprint
  {http://arxiv.org/abs/1712.05815} {arXiv:1712.05815 [nucl-th]} \BibitemShut
  {NoStop}%
\bibitem [{\citenamefont {Bilandzic}\ \emph {et~al.}(2011)\citenamefont
  {Bilandzic}, \citenamefont {Snellings},\ and\ \citenamefont
  {Voloshin}}]{Bilandzic:2010jr}%
  \BibitemOpen
  \bibfield  {author} {\bibinfo {author} {\bibfnamefont {Ante}\ \bibnamefont
  {Bilandzic}}, \bibinfo {author} {\bibfnamefont {Raimond}\ \bibnamefont
  {Snellings}}, \ and\ \bibinfo {author} {\bibfnamefont {Sergei}\ \bibnamefont
  {Voloshin}},\ }\bibfield  {title} {\enquote {\bibinfo {title} {{Flow analysis
  with cumulants: Direct calculations}},}\ }\href {\doibase
  10.1103/PhysRevC.83.044913} {\bibfield  {journal} {\bibinfo  {journal} {Phys.
  Rev. C}\ }\textbf {\bibinfo {volume} {83}},\ \bibinfo {pages} {044913}
  (\bibinfo {year} {2011})},\ \Eprint {http://arxiv.org/abs/1010.0233}
  {arXiv:1010.0233 [nucl-ex]} \BibitemShut {NoStop}%
\bibitem [{\citenamefont {Bilandzic}\ \emph {et~al.}(2014)\citenamefont
  {Bilandzic}, \citenamefont {Christensen}, \citenamefont {Gulbrandsen},
  \citenamefont {Hansen},\ and\ \citenamefont {Zhou}}]{Bilandzic:2013kga}%
  \BibitemOpen
  \bibfield  {author} {\bibinfo {author} {\bibfnamefont {Ante}\ \bibnamefont
  {Bilandzic}}, \bibinfo {author} {\bibfnamefont {Christian~Holm}\ \bibnamefont
  {Christensen}}, \bibinfo {author} {\bibfnamefont {Kristjan}\ \bibnamefont
  {Gulbrandsen}}, \bibinfo {author} {\bibfnamefont {Alexander}\ \bibnamefont
  {Hansen}}, \ and\ \bibinfo {author} {\bibfnamefont {You}\ \bibnamefont
  {Zhou}},\ }\bibfield  {title} {\enquote {\bibinfo {title} {{Generic framework
  for anisotropic flow analyses with multiparticle azimuthal correlations}},}\
  }\href {\doibase 10.1103/PhysRevC.89.064904} {\bibfield  {journal} {\bibinfo
  {journal} {Phys. Rev. C}\ }\textbf {\bibinfo {volume} {89}},\ \bibinfo
  {pages} {064904} (\bibinfo {year} {2014})},\ \Eprint
  {http://arxiv.org/abs/1312.3572} {arXiv:1312.3572 [nucl-ex]} \BibitemShut
  {NoStop}%
\bibitem [{\citenamefont {Jia}\ \emph {et~al.}(2017)\citenamefont {Jia},
  \citenamefont {Zhou},\ and\ \citenamefont {Trzupek}}]{Jia:2017hbm}%
  \BibitemOpen
  \bibfield  {author} {\bibinfo {author} {\bibfnamefont {Jiangyong}\
  \bibnamefont {Jia}}, \bibinfo {author} {\bibfnamefont {Mingliang}\
  \bibnamefont {Zhou}}, \ and\ \bibinfo {author} {\bibfnamefont {Adam}\
  \bibnamefont {Trzupek}},\ }\bibfield  {title} {\enquote {\bibinfo {title}
  {{Revealing long-range multiparticle collectivity in small collision systems
  via subevent cumulants}},}\ }\href {\doibase 10.1103/PhysRevC.96.034906}
  {\bibfield  {journal} {\bibinfo  {journal} {Phys. Rev. C}\ }\textbf {\bibinfo
  {volume} {96}},\ \bibinfo {pages} {034906} (\bibinfo {year} {2017})},\
  \Eprint {http://arxiv.org/abs/1701.03830} {arXiv:1701.03830 [nucl-th]}
  \BibitemShut {NoStop}%
\bibitem [{\citenamefont {Abdallah}\ \emph {et~al.}(2022)\citenamefont
  {Abdallah} \emph {et~al.}}]{STAR:2021mii}%
  \BibitemOpen
  \bibfield  {author} {\bibinfo {author} {\bibfnamefont {Mohamed}\ \bibnamefont
  {Abdallah}} \emph {et~al.} (\bibinfo {collaboration} {STAR}),\ }\bibfield
  {title} {\enquote {\bibinfo {title} {{Search for the chiral magnetic effect
  with isobar collisions at $\sqrt {s_{NN}}$=200 GeV by the STAR Collaboration
  at the BNL Relativistic Heavy Ion Collider}},}\ }\href {\doibase
  10.1103/PhysRevC.105.014901} {\bibfield  {journal} {\bibinfo  {journal}
  {Phys. Rev. C}\ }\textbf {\bibinfo {volume} {105}},\ \bibinfo {pages}
  {014901} (\bibinfo {year} {2022})},\ \Eprint
  {http://arxiv.org/abs/2109.00131} {arXiv:2109.00131 [nucl-ex]} \BibitemShut
  {NoStop}%
\bibitem [{\citenamefont {Zhang}\ and\ \citenamefont
  {Jia}(2022)}]{Zhang:2021kxj}%
  \BibitemOpen
  \bibfield  {author} {\bibinfo {author} {\bibfnamefont {Chunjian}\
  \bibnamefont {Zhang}}\ and\ \bibinfo {author} {\bibfnamefont {Jiangyong}\
  \bibnamefont {Jia}},\ }\bibfield  {title} {\enquote {\bibinfo {title}
  {{Evidence of Quadrupole and Octupole Deformations in Zr96+Zr96 and Ru96+Ru96
  Collisions at Ultrarelativistic Energies}},}\ }\href {\doibase
  10.1103/PhysRevLett.128.022301} {\bibfield  {journal} {\bibinfo  {journal}
  {Phys. Rev. Lett.}\ }\textbf {\bibinfo {volume} {128}},\ \bibinfo {pages}
  {022301} (\bibinfo {year} {2022})},\ \Eprint
  {http://arxiv.org/abs/2109.01631} {arXiv:2109.01631 [nucl-th]} \BibitemShut
  {NoStop}%
\bibitem [{\citenamefont {Xu}\ \emph {et~al.}(2021)\citenamefont {Xu},
  \citenamefont {Zhao}, \citenamefont {Li}, \citenamefont {Zhou}, \citenamefont
  {Chen},\ and\ \citenamefont {Wang}}]{Xu:2021uar}%
  \BibitemOpen
  \bibfield  {author} {\bibinfo {author} {\bibfnamefont {Hao-jie}\ \bibnamefont
  {Xu}}, \bibinfo {author} {\bibfnamefont {Wenbin}\ \bibnamefont {Zhao}},
  \bibinfo {author} {\bibfnamefont {Hanlin}\ \bibnamefont {Li}}, \bibinfo
  {author} {\bibfnamefont {Ying}\ \bibnamefont {Zhou}}, \bibinfo {author}
  {\bibfnamefont {Lie-Wen}\ \bibnamefont {Chen}}, \ and\ \bibinfo {author}
  {\bibfnamefont {Fuqiang}\ \bibnamefont {Wang}},\ }\bibfield  {title}
  {\enquote {\bibinfo {title} {{Probing nuclear structure with mean transverse
  momentum in relativistic isobar collisions}},}\ }\href@noop {} {\  (\bibinfo
  {year} {2021})},\ \Eprint {http://arxiv.org/abs/2111.14812} {arXiv:2111.14812
  [nucl-th]} \BibitemShut {NoStop}%
\bibitem [{\citenamefont {Liu}\ \emph {et~al.}(2022)\citenamefont {Liu},
  \citenamefont {Zhang}, \citenamefont {Zhou}, \citenamefont {Xu},
  \citenamefont {Jia},\ and\ \citenamefont {Peng}}]{Liu:2022kvz}%
  \BibitemOpen
  \bibfield  {author} {\bibinfo {author} {\bibfnamefont {Lu-Meng}\ \bibnamefont
  {Liu}}, \bibinfo {author} {\bibfnamefont {Chun-Jian}\ \bibnamefont {Zhang}},
  \bibinfo {author} {\bibfnamefont {Jia}\ \bibnamefont {Zhou}}, \bibinfo
  {author} {\bibfnamefont {Jun}\ \bibnamefont {Xu}}, \bibinfo {author}
  {\bibfnamefont {Jiangyong}\ \bibnamefont {Jia}}, \ and\ \bibinfo {author}
  {\bibfnamefont {Guang-Xiong}\ \bibnamefont {Peng}},\ }\bibfield  {title}
  {\enquote {\bibinfo {title} {{Probing neutron-skin thickness with free
  spectator neutrons in ultracentral high-energy isobaric collisions}},}\
  }\href {\doibase 10.1016/j.physletb.2022.137441} {\bibfield  {journal}
  {\bibinfo  {journal} {Phys. Lett. B}\ }\textbf {\bibinfo {volume} {834}},\
  \bibinfo {pages} {137441} (\bibinfo {year} {2022})},\ \Eprint
  {http://arxiv.org/abs/2203.09924} {arXiv:2203.09924 [nucl-th]} \BibitemShut
  {NoStop}%
\bibitem [{\citenamefont {Zhao}\ \emph {et~al.}(2023)\citenamefont {Zhao},
  \citenamefont {Xu}, \citenamefont {Liu},\ and\ \citenamefont
  {Song}}]{Zhao:2022uhl}%
  \BibitemOpen
  \bibfield  {author} {\bibinfo {author} {\bibfnamefont {Shujun}\ \bibnamefont
  {Zhao}}, \bibinfo {author} {\bibfnamefont {Hao-jie}\ \bibnamefont {Xu}},
  \bibinfo {author} {\bibfnamefont {Yu-Xin}\ \bibnamefont {Liu}}, \ and\
  \bibinfo {author} {\bibfnamefont {Huichao}\ \bibnamefont {Song}},\ }\bibfield
   {title} {\enquote {\bibinfo {title} {{Probing the nuclear deformation with
  three-particle asymmetric cumulant in RHIC isobar runs}},}\ }\href {\doibase
  10.1016/j.physletb.2023.137838} {\bibfield  {journal} {\bibinfo  {journal}
  {Phys. Lett. B}\ }\textbf {\bibinfo {volume} {839}},\ \bibinfo {pages}
  {137838} (\bibinfo {year} {2023})},\ \Eprint
  {http://arxiv.org/abs/2204.02387} {arXiv:2204.02387 [nucl-th]} \BibitemShut
  {NoStop}%
\bibitem [{\citenamefont {Jia}\ \emph {et~al.}(2023)\citenamefont {Jia},
  \citenamefont {Giacalone},\ and\ \citenamefont {Zhang}}]{Jia:2022qrq}%
  \BibitemOpen
  \bibfield  {author} {\bibinfo {author} {\bibfnamefont {Jiangyong}\
  \bibnamefont {Jia}}, \bibinfo {author} {\bibfnamefont {Giuliano}\
  \bibnamefont {Giacalone}}, \ and\ \bibinfo {author} {\bibfnamefont
  {Chunjian}\ \bibnamefont {Zhang}},\ }\bibfield  {title} {\enquote {\bibinfo
  {title} {{Precision Tests of the Nonlinear Mode Coupling of Anisotropic Flow
  via High-Energy Collisions of Isobars}},}\ }\href {\doibase
  10.1088/0256-307X/40/4/042501} {\bibfield  {journal} {\bibinfo  {journal}
  {Chin. Phys. Lett.}\ }\textbf {\bibinfo {volume} {40}},\ \bibinfo {pages}
  {042501} (\bibinfo {year} {2023})},\ \Eprint
  {http://arxiv.org/abs/2206.07184} {arXiv:2206.07184 [nucl-th]} \BibitemShut
  {NoStop}%
\bibitem [{\citenamefont {Jia}\ and\ \citenamefont
  {Zhang}(2023)}]{Jia:2021oyt}%
  \BibitemOpen
  \bibfield  {author} {\bibinfo {author} {\bibfnamefont {Jiangyong}\
  \bibnamefont {Jia}}\ and\ \bibinfo {author} {\bibfnamefont {Chunjian}\
  \bibnamefont {Zhang}},\ }\bibfield  {title} {\enquote {\bibinfo {title}
  {{Scaling approach to nuclear structure in high-energy heavy-ion
  collisions}},}\ }\href {\doibase 10.1103/PhysRevC.107.L021901} {\bibfield
  {journal} {\bibinfo  {journal} {Phys. Rev. C}\ }\textbf {\bibinfo {volume}
  {107}},\ \bibinfo {pages} {L021901} (\bibinfo {year} {2023})},\ \Eprint
  {http://arxiv.org/abs/2111.15559} {arXiv:2111.15559 [nucl-th]} \BibitemShut
  {NoStop}%
\bibitem [{\citenamefont {Voloshin}\ \emph {et~al.}(2008)\citenamefont
  {Voloshin}, \citenamefont {Poskanzer}, \citenamefont {Tang},\ and\
  \citenamefont {Wang}}]{Voloshin:2007pc}%
  \BibitemOpen
  \bibfield  {author} {\bibinfo {author} {\bibfnamefont {Sergei~A.}\
  \bibnamefont {Voloshin}}, \bibinfo {author} {\bibfnamefont {Arthur~M.}\
  \bibnamefont {Poskanzer}}, \bibinfo {author} {\bibfnamefont {Aihong}\
  \bibnamefont {Tang}}, \ and\ \bibinfo {author} {\bibfnamefont {Gang}\
  \bibnamefont {Wang}},\ }\bibfield  {title} {\enquote {\bibinfo {title}
  {{Elliptic flow in the Gaussian model of eccentricity fluctuations}},}\
  }\href {\doibase 10.1016/j.physletb.2007.11.043} {\bibfield  {journal}
  {\bibinfo  {journal} {Phys. Lett. B}\ }\textbf {\bibinfo {volume} {659}},\
  \bibinfo {pages} {537--541} (\bibinfo {year} {2008})},\ \Eprint
  {http://arxiv.org/abs/0708.0800} {arXiv:0708.0800 [nucl-th]} \BibitemShut
  {NoStop}%
\bibitem [{Note1()}]{Note1}%
  \BibitemOpen
  \bibinfo {note} {Note that changes in nuclear structure also affect the
  distribution $p(N_{\protect \mathrm {part}})$, such that the events with the
  same $N_{\protect \mathrm {part}}$ correspond to slightly different
  centralities and vice versa~\cite {Jia:2022iji}. This secondary effect
  introduces a small correlation between $\varepsilon _2^{\protect \mathrm
  {rp}}$ and $\beta _n$, but it is subleading compared to the one discussed in
  Fig.~\ref {fig:1}.}\BibitemShut {Stop}%
\bibitem [{\citenamefont {Lin}\ \emph {et~al.}(2005)\citenamefont {Lin},
  \citenamefont {Ko}, \citenamefont {Li}, \citenamefont {Zhang},\ and\
  \citenamefont {Pal}}]{Lin:2004en}%
  \BibitemOpen
  \bibfield  {author} {\bibinfo {author} {\bibfnamefont {Zi-Wei}\ \bibnamefont
  {Lin}}, \bibinfo {author} {\bibfnamefont {Che~Ming}\ \bibnamefont {Ko}},
  \bibinfo {author} {\bibfnamefont {Bao-An}\ \bibnamefont {Li}}, \bibinfo
  {author} {\bibfnamefont {Bin}\ \bibnamefont {Zhang}}, \ and\ \bibinfo
  {author} {\bibfnamefont {Subrata}\ \bibnamefont {Pal}},\ }\bibfield  {title}
  {\enquote {\bibinfo {title} {{A Multi-phase transport model for relativistic
  heavy ion collisions}},}\ }\href {\doibase 10.1103/PhysRevC.72.064901}
  {\bibfield  {journal} {\bibinfo  {journal} {Phys. Rev.}\ }\textbf {\bibinfo
  {volume} {C72}},\ \bibinfo {pages} {064901} (\bibinfo {year} {2005})},\
  \Eprint {http://arxiv.org/abs/nucl-th/0411110} {arXiv:nucl-th/0411110
  [nucl-th]} \BibitemShut {NoStop}%
\bibitem [{\citenamefont {Ma}\ and\ \citenamefont {Bzdak}(2014)}]{Ma:2014pva}%
  \BibitemOpen
  \bibfield  {author} {\bibinfo {author} {\bibfnamefont {Guo-Liang}\
  \bibnamefont {Ma}}\ and\ \bibinfo {author} {\bibfnamefont {Adam}\
  \bibnamefont {Bzdak}},\ }\bibfield  {title} {\enquote {\bibinfo {title}
  {{Long-range azimuthal correlations in proton–proton and proton–nucleus
  collisions from the incoherent scattering of partons}},}\ }\href {\doibase
  10.1016/j.physletb.2014.10.066} {\bibfield  {journal} {\bibinfo  {journal}
  {Phys. Lett.}\ }\textbf {\bibinfo {volume} {B739}},\ \bibinfo {pages}
  {209--213} (\bibinfo {year} {2014})},\ \Eprint
  {http://arxiv.org/abs/1404.4129} {arXiv:1404.4129 [hep-ph]} \BibitemShut
  {NoStop}%
\bibitem [{\citenamefont {Bzdak}\ and\ \citenamefont
  {Ma}(2014)}]{Bzdak:2014dia}%
  \BibitemOpen
  \bibfield  {author} {\bibinfo {author} {\bibfnamefont {Adam}\ \bibnamefont
  {Bzdak}}\ and\ \bibinfo {author} {\bibfnamefont {Guo-Liang}\ \bibnamefont
  {Ma}},\ }\bibfield  {title} {\enquote {\bibinfo {title} {{Elliptic and
  triangular flow in $p$+Pb and peripheral Pb+Pb collisions from parton
  scatterings}},}\ }\href {\doibase 10.1103/PhysRevLett.113.252301} {\bibfield
  {journal} {\bibinfo  {journal} {Phys. Rev. Lett.}\ }\textbf {\bibinfo
  {volume} {113}},\ \bibinfo {pages} {252301} (\bibinfo {year} {2014})},\
  \Eprint {http://arxiv.org/abs/1406.2804} {arXiv:1406.2804 [hep-ph]}
  \BibitemShut {NoStop}%
\bibitem [{\citenamefont {Giacalone}\ \emph
  {et~al.}(2017{\natexlab{a}})\citenamefont {Giacalone}, \citenamefont
  {Noronha-Hostler},\ and\ \citenamefont {Ollitrault}}]{Giacalone:2017uqx}%
  \BibitemOpen
  \bibfield  {author} {\bibinfo {author} {\bibfnamefont {Giuliano}\
  \bibnamefont {Giacalone}}, \bibinfo {author} {\bibfnamefont {Jacquelyn}\
  \bibnamefont {Noronha-Hostler}}, \ and\ \bibinfo {author} {\bibfnamefont
  {Jean-Yves}\ \bibnamefont {Ollitrault}},\ }\bibfield  {title} {\enquote
  {\bibinfo {title} {{Relative flow fluctuations as a probe of initial state
  fluctuations}},}\ }\href {\doibase 10.1103/PhysRevC.95.054910} {\bibfield
  {journal} {\bibinfo  {journal} {Phys. Rev. C}\ }\textbf {\bibinfo {volume}
  {95}},\ \bibinfo {pages} {054910} (\bibinfo {year} {2017}{\natexlab{a}})},\
  \Eprint {http://arxiv.org/abs/1702.01730} {arXiv:1702.01730 [nucl-th]}
  \BibitemShut {NoStop}%
\bibitem [{\citenamefont {Nijs}\ and\ \citenamefont {van~der
  Schee}(2021)}]{Nijs:2021kvn}%
  \BibitemOpen
  \bibfield  {author} {\bibinfo {author} {\bibfnamefont {Govert}\ \bibnamefont
  {Nijs}}\ and\ \bibinfo {author} {\bibfnamefont {Wilke}\ \bibnamefont {van~der
  Schee}},\ }\bibfield  {title} {\enquote {\bibinfo {title} {{Inferring nuclear
  structure from heavy isobar collisions using Trajectum}},}\ }\href@noop {} {\
   (\bibinfo {year} {2021})},\ \Eprint {http://arxiv.org/abs/2112.13771}
  {arXiv:2112.13771 [nucl-th]} \BibitemShut {NoStop}%
\bibitem [{\citenamefont {{ALICE Collaboration}}(2022)}]{ALICE:2022xhd}%
  \BibitemOpen
  \bibfield  {author} {\bibinfo {author} {\bibnamefont {{ALICE
  Collaboration}}},\ }\bibfield  {title} {\enquote {\bibinfo {title} {{Elliptic
  flow of charged particles at midrapidity relative to the spectator plane in
  Pb-Pb and Xe-Xe collisions}},}\ }\href@noop {} {\  (\bibinfo {year}
  {2022})},\ \Eprint {http://arxiv.org/abs/2204.10240} {arXiv:2204.10240
  [nucl-ex]} \BibitemShut {NoStop}%
\bibitem [{\citenamefont {Teaney}\ and\ \citenamefont
  {Yan}(2011)}]{Teaney:2010vd}%
  \BibitemOpen
  \bibfield  {author} {\bibinfo {author} {\bibfnamefont {Derek}\ \bibnamefont
  {Teaney}}\ and\ \bibinfo {author} {\bibfnamefont {Li}~\bibnamefont {Yan}},\
  }\bibfield  {title} {\enquote {\bibinfo {title} {{Triangularity and Dipole
  Asymmetry in Heavy Ion Collisions}},}\ }\href {\doibase
  10.1103/PhysRevC.83.064904} {\bibfield  {journal} {\bibinfo  {journal} {Phys.
  Rev. C}\ }\textbf {\bibinfo {volume} {83}},\ \bibinfo {pages} {064904}
  (\bibinfo {year} {2011})},\ \Eprint {http://arxiv.org/abs/1010.1876}
  {arXiv:1010.1876 [nucl-th]} \BibitemShut {NoStop}%
\bibitem [{\citenamefont {Giacalone}\ \emph
  {et~al.}(2017{\natexlab{b}})\citenamefont {Giacalone}, \citenamefont {Yan},
  \citenamefont {Noronha-Hostler},\ and\ \citenamefont
  {Ollitrault}}]{Giacalone:2016eyu}%
  \BibitemOpen
  \bibfield  {author} {\bibinfo {author} {\bibfnamefont {Giuliano}\
  \bibnamefont {Giacalone}}, \bibinfo {author} {\bibfnamefont {Li}~\bibnamefont
  {Yan}}, \bibinfo {author} {\bibfnamefont {Jacquelyn}\ \bibnamefont
  {Noronha-Hostler}}, \ and\ \bibinfo {author} {\bibfnamefont {Jean-Yves}\
  \bibnamefont {Ollitrault}},\ }\bibfield  {title} {\enquote {\bibinfo {title}
  {{Skewness of elliptic flow fluctuations}},}\ }\href {\doibase
  10.1103/PhysRevC.95.014913} {\bibfield  {journal} {\bibinfo  {journal} {Phys.
  Rev. C}\ }\textbf {\bibinfo {volume} {95}},\ \bibinfo {pages} {014913}
  (\bibinfo {year} {2017}{\natexlab{b}})},\ \Eprint
  {http://arxiv.org/abs/1608.01823} {arXiv:1608.01823 [nucl-th]} \BibitemShut
  {NoStop}%
\bibitem [{\citenamefont {Bhalerao}\ \emph {et~al.}(2019)\citenamefont
  {Bhalerao}, \citenamefont {Giacalone},\ and\ \citenamefont
  {Ollitrault}}]{Bhalerao:2018anl}%
  \BibitemOpen
  \bibfield  {author} {\bibinfo {author} {\bibfnamefont {Rajeev~S.}\
  \bibnamefont {Bhalerao}}, \bibinfo {author} {\bibfnamefont {Giuliano}\
  \bibnamefont {Giacalone}}, \ and\ \bibinfo {author} {\bibfnamefont
  {Jean-Yves}\ \bibnamefont {Ollitrault}},\ }\bibfield  {title} {\enquote
  {\bibinfo {title} {{Kurtosis of elliptic flow fluctuations}},}\ }\href
  {\doibase 10.1103/PhysRevC.99.014907} {\bibfield  {journal} {\bibinfo
  {journal} {Phys. Rev. C}\ }\textbf {\bibinfo {volume} {99}},\ \bibinfo
  {pages} {014907} (\bibinfo {year} {2019})},\ \Eprint
  {http://arxiv.org/abs/1811.00837} {arXiv:1811.00837 [nucl-th]} \BibitemShut
  {NoStop}%
\bibitem [{\citenamefont {Giacalone}(2019)}]{Giacalone:2018apa}%
  \BibitemOpen
  \bibfield  {author} {\bibinfo {author} {\bibfnamefont {Giuliano}\
  \bibnamefont {Giacalone}},\ }\bibfield  {title} {\enquote {\bibinfo {title}
  {{Elliptic flow fluctuations in central collisions of spherical and deformed
  nuclei}},}\ }\href {\doibase 10.1103/PhysRevC.99.024910} {\bibfield
  {journal} {\bibinfo  {journal} {Phys. Rev. C}\ }\textbf {\bibinfo {volume}
  {99}},\ \bibinfo {pages} {024910} (\bibinfo {year} {2019})},\ \Eprint
  {http://arxiv.org/abs/1811.03959} {arXiv:1811.03959 [nucl-th]} \BibitemShut
  {NoStop}%
\bibitem [{\citenamefont {Everett}\ \emph {et~al.}(2021)\citenamefont {Everett}
  \emph {et~al.}}]{JETSCAPE:2020mzn}%
  \BibitemOpen
  \bibfield  {author} {\bibinfo {author} {\bibfnamefont {D.}~\bibnamefont
  {Everett}} \emph {et~al.} (\bibinfo {collaboration} {JETSCAPE}),\ }\bibfield
  {title} {\enquote {\bibinfo {title} {{Multisystem Bayesian constraints on the
  transport coefficients of QCD matter}},}\ }\href {\doibase
  10.1103/PhysRevC.103.054904} {\bibfield  {journal} {\bibinfo  {journal}
  {Phys. Rev. C}\ }\textbf {\bibinfo {volume} {103}},\ \bibinfo {pages}
  {054904} (\bibinfo {year} {2021})},\ \Eprint
  {http://arxiv.org/abs/2011.01430} {arXiv:2011.01430 [hep-ph]} \BibitemShut
  {NoStop}%
\bibitem [{\citenamefont {Nijs}\ \emph {et~al.}(2021)\citenamefont {Nijs},
  \citenamefont {van~der Schee}, \citenamefont {G\"ursoy},\ and\ \citenamefont
  {Snellings}}]{Nijs:2020ors}%
  \BibitemOpen
  \bibfield  {author} {\bibinfo {author} {\bibfnamefont {Govert}\ \bibnamefont
  {Nijs}}, \bibinfo {author} {\bibfnamefont {Wilke}\ \bibnamefont {van~der
  Schee}}, \bibinfo {author} {\bibfnamefont {Umut}\ \bibnamefont {G\"ursoy}}, \
  and\ \bibinfo {author} {\bibfnamefont {Raimond}\ \bibnamefont {Snellings}},\
  }\bibfield  {title} {\enquote {\bibinfo {title} {{Transverse Momentum
  Differential Global Analysis of Heavy-Ion Collisions}},}\ }\href {\doibase
  10.1103/PhysRevLett.126.202301} {\bibfield  {journal} {\bibinfo  {journal}
  {Phys. Rev. Lett.}\ }\textbf {\bibinfo {volume} {126}},\ \bibinfo {pages}
  {202301} (\bibinfo {year} {2021})},\ \Eprint
  {http://arxiv.org/abs/2010.15130} {arXiv:2010.15130 [nucl-th]} \BibitemShut
  {NoStop}%
\bibitem [{\citenamefont {Bally}\ \emph
  {et~al.}(2022{\natexlab{b}})\citenamefont {Bally} \emph
  {et~al.}}]{Bally:2022vgo}%
  \BibitemOpen
  \bibfield  {author} {\bibinfo {author} {\bibfnamefont {Benjamin}\
  \bibnamefont {Bally}} \emph {et~al.},\ }\bibfield  {title} {\enquote
  {\bibinfo {title} {{Imaging the initial condition of heavy-ion collisions and
  nuclear structure across the nuclide chart}},}\ }\href@noop {} {\  (\bibinfo
  {year} {2022}{\natexlab{b}})},\ \Eprint {http://arxiv.org/abs/2209.11042}
  {arXiv:2209.11042 [nucl-ex]} \BibitemShut {NoStop}%
\bibitem [{\citenamefont {Citron}\ \emph {et~al.}(2019)\citenamefont {Citron}
  \emph {et~al.}}]{Citron:2018lsq}%
  \BibitemOpen
  \bibfield  {author} {\bibinfo {author} {\bibfnamefont {Z.}~\bibnamefont
  {Citron}} \emph {et~al.},\ }\bibfield  {title} {\enquote {\bibinfo {title}
  {{Report from Working Group 5}: {Future physics opportunities for
  high-density QCD at the LHC with heavy-ion and proton beams}},}\ }\href
  {\doibase 10.23731/CYRM-2019-007.1159} {\bibfield  {journal} {\bibinfo
  {journal} {CERN Yellow Rep. Monogr.}\ }\textbf {\bibinfo {volume} {7}},\
  \bibinfo {pages} {1159--1410} (\bibinfo {year} {2019})},\ \Eprint
  {http://arxiv.org/abs/1812.06772} {arXiv:1812.06772 [hep-ph]} \BibitemShut
  {NoStop}%
\bibitem [{\citenamefont {Miller}\ \emph {et~al.}(2007)\citenamefont {Miller},
  \citenamefont {Reygers}, \citenamefont {Sanders},\ and\ \citenamefont
  {Steinberg}}]{Miller:2007ri}%
  \BibitemOpen
  \bibfield  {author} {\bibinfo {author} {\bibfnamefont {Michael~L.}\
  \bibnamefont {Miller}}, \bibinfo {author} {\bibfnamefont {Klaus}\
  \bibnamefont {Reygers}}, \bibinfo {author} {\bibfnamefont {Stephen~J.}\
  \bibnamefont {Sanders}}, \ and\ \bibinfo {author} {\bibfnamefont {Peter}\
  \bibnamefont {Steinberg}},\ }\bibfield  {title} {\enquote {\bibinfo {title}
  {{Glauber modeling in high energy nuclear collisions}},}\ }\href {\doibase
  10.1146/annurev.nucl.57.090506.123020} {\bibfield  {journal} {\bibinfo
  {journal} {Ann. Rev. Nucl. Part. Sci.}\ }\textbf {\bibinfo {volume} {57}},\
  \bibinfo {pages} {205--243} (\bibinfo {year} {2007})},\ \Eprint
  {http://arxiv.org/abs/nucl-ex/0701025} {arXiv:nucl-ex/0701025} \BibitemShut
  {NoStop}%
\bibitem [{\citenamefont {Aad}\ \emph {et~al.}(2013)\citenamefont {Aad} \emph
  {et~al.}}]{ATLAS:2013xzf}%
  \BibitemOpen
  \bibfield  {author} {\bibinfo {author} {\bibfnamefont {Georges}\ \bibnamefont
  {Aad}} \emph {et~al.} (\bibinfo {collaboration} {ATLAS}),\ }\bibfield
  {title} {\enquote {\bibinfo {title} {{Measurement of the distributions of
  event-by-event flow harmonics in lead-lead collisions at = 2.76 TeV with the
  ATLAS detector at the LHC}},}\ }\href {\doibase 10.1007/JHEP11(2013)183}
  {\bibfield  {journal} {\bibinfo  {journal} {JHEP}\ }\textbf {\bibinfo
  {volume} {11}},\ \bibinfo {pages} {183} (\bibinfo {year} {2013})},\ \Eprint
  {http://arxiv.org/abs/1305.2942} {arXiv:1305.2942 [hep-ex]} \BibitemShut
  {NoStop}%
\bibitem [{\citenamefont {Sirunyan}\ \emph {et~al.}(2019)\citenamefont
  {Sirunyan} \emph {et~al.}}]{CMS:2017glf}%
  \BibitemOpen
  \bibfield  {author} {\bibinfo {author} {\bibfnamefont {Albert~M}\
  \bibnamefont {Sirunyan}} \emph {et~al.} (\bibinfo {collaboration} {CMS}),\
  }\bibfield  {title} {\enquote {\bibinfo {title} {{Non-Gaussian elliptic-flow
  fluctuations in PbPb collisions at $\sqrt{\smash[b]{s_{_\text{NN}}}} = 5.02$
  TeV}},}\ }\href {\doibase 10.1016/j.physletb.2018.11.063} {\bibfield
  {journal} {\bibinfo  {journal} {Phys. Lett. B}\ }\textbf {\bibinfo {volume}
  {789}},\ \bibinfo {pages} {643--665} (\bibinfo {year} {2019})},\ \Eprint
  {http://arxiv.org/abs/1711.05594} {arXiv:1711.05594 [nucl-ex]} \BibitemShut
  {NoStop}%
\bibitem [{\citenamefont {Acharya}\ \emph {et~al.}(2018)\citenamefont {Acharya}
  \emph {et~al.}}]{ALICE:2018rtz}%
  \BibitemOpen
  \bibfield  {author} {\bibinfo {author} {\bibfnamefont {S.}~\bibnamefont
  {Acharya}} \emph {et~al.} (\bibinfo {collaboration} {ALICE}),\ }\bibfield
  {title} {\enquote {\bibinfo {title} {{Energy dependence and fluctuations of
  anisotropic flow in Pb-Pb collisions at $ \sqrt{s_{\mathrm{NN}}}=5.02 $ and
  2.76 TeV}},}\ }\href {\doibase 10.1007/JHEP07(2018)103} {\bibfield  {journal}
  {\bibinfo  {journal} {JHEP}\ }\textbf {\bibinfo {volume} {07}},\ \bibinfo
  {pages} {103} (\bibinfo {year} {2018})},\ \Eprint
  {http://arxiv.org/abs/1804.02944} {arXiv:1804.02944 [nucl-ex]} \BibitemShut
  {NoStop}%
\bibitem [{\citenamefont {Aaboud}\ \emph {et~al.}(2020)\citenamefont {Aaboud}
  \emph {et~al.}}]{ATLAS:2019peb}%
  \BibitemOpen
  \bibfield  {author} {\bibinfo {author} {\bibfnamefont {Morad}\ \bibnamefont
  {Aaboud}} \emph {et~al.} (\bibinfo {collaboration} {ATLAS}),\ }\bibfield
  {title} {\enquote {\bibinfo {title} {{Fluctuations of anisotropic flow in
  Pb+Pb collisions at $ \sqrt{{\mathrm{s}}_{\mathrm{NN}}} $ = 5.02 TeV with the
  ATLAS detector}},}\ }\href {\doibase 10.1007/JHEP01(2020)051} {\bibfield
  {journal} {\bibinfo  {journal} {JHEP}\ }\textbf {\bibinfo {volume} {01}},\
  \bibinfo {pages} {051} (\bibinfo {year} {2020})},\ \Eprint
  {http://arxiv.org/abs/1904.04808} {arXiv:1904.04808 [nucl-ex]} \BibitemShut
  {NoStop}%
\bibitem [{\citenamefont {Jia}\ \emph {et~al.}(2022{\natexlab{b}})\citenamefont
  {Jia}, \citenamefont {Wang},\ and\ \citenamefont {Zhang}}]{Jia:2022iji}%
  \BibitemOpen
  \bibfield  {author} {\bibinfo {author} {\bibfnamefont {Jiangyong}\
  \bibnamefont {Jia}}, \bibinfo {author} {\bibfnamefont {Gang}\ \bibnamefont
  {Wang}}, \ and\ \bibinfo {author} {\bibfnamefont {Chunjian}\ \bibnamefont
  {Zhang}},\ }\bibfield  {title} {\enquote {\bibinfo {title} {{Impact of event
  activity variable on the ratio observables in isobar collisions}},}\ }\href
  {\doibase 10.1016/j.physletb.2022.137312} {\bibfield  {journal} {\bibinfo
  {journal} {Phys. Lett. B}\ }\textbf {\bibinfo {volume} {833}},\ \bibinfo
  {pages} {137312} (\bibinfo {year} {2022}{\natexlab{b}})},\ \Eprint
  {http://arxiv.org/abs/2203.12654} {arXiv:2203.12654 [nucl-th]} \BibitemShut
  {NoStop}%
\end{thebibliography}%
\end{document}